\documentclass[journal,twoside,web]{ieeecolor}
\usepackage{generic}
\usepackage{cite}
\usepackage{amsmath,amssymb,amsfonts}
\usepackage{graphicx}
\usepackage{textcomp}
\usepackage{mathrsfs}
\usepackage{geometry}
\usepackage{multicol} %更多栏
\usepackage{wrapfig}
\usepackage{soul}%高亮文本
\usepackage{etoolbox}
\usepackage{tikz}

\makeatletter
\let\NAT@parse\undefined
\makeatother

\AddToHook{cmd/label/before}{\phantomsection}

\usepackage[hidelinks=true]{hyperref}

\usetikzlibrary{shapes.geometric, arrows}
\usetikzlibrary{arrows.meta, positioning, calc}
\tikzstyle{circleNode} = [circle, minimum size=0.8cm, text centered, draw=black]
\tikzstyle{arrow} = [thick,->,>=stealth]
\makeatletter

\newenvironment{proof}[1][Proof]{%
  \par% 
  \topsep6\p@\@plus6\p@\relax%
  \trivlist%
  \setlength{\leftmargin}{2em}%
  \setlength{\labelsep}{0.5em}% 
  \item[\hskip\labelsep\itshape #1:]%
  \ignorespaces
}{%
  \hspace*{\fill}~\QED
  \endtrivlist
  \@endpefalse
}
\makeatother

\makeatletter
\def\@begintheorem#1#2{
  \par% 
  \vspace{4pt plus 1pt minus 1pt}
  \trivlist
  \item[\hskip\labelsep\bfseries #1\ #2.]
  \noindent%
  \ignorespaces
}
\def\@endtheorem{%
  \endtrivlist
  \vspace{4pt plus 1pt minus 1pt}
  \@endpefalse
}
\makeatother

\renewcommand{\epsilon}{\varepsilon}

\makeatletter%下4行为罗马字母

\newcommand{\plat}[1]{\raisebox{0pt}[0pt][0pt]{#1}}

\newcommand{\Rmnum}[1]{\expandafter\@slowromancap\romannumeral #1@}
\makeatother
\newcommand{\powerset}{\raisebox{.15\baselineskip}{\Large\ensuremath{\wp}}}

\newtheorem{theorem}{Theorem}
\newtheorem{lemma}{Lemma}

\newtheorem{observation}{Observation}
\newtheorem{example}{Example}

\newtheorem{proposition}{Proposition}
\newtheorem{corollary}{Corollary}

\newtheorem{myDef}{Definition}

\newcommand{\SP}{\textit{SP}}
\newcommand{\MSP}{\textit{MSP}}

 \BeforeBeginEnvironment{proof}{\vspace{-0.5\baselineskip}}
 \AfterEndEnvironment{proof}{\vspace{-0.5\baselineskip}}

\begin{document}
\title{More on Maximally Permissive Similarity Control of Discrete Event Systems}
\author{Yu Wang, Zhaohui Zhu$^*$, Rob van Glabbeek, Shigemasa Takai, Jinjin Zhang, and Lixing Tan
\thanks{Yu Wang is with the College of Computer Science and Technology, Nanjing University of Aeronautics and Astronautics, Nanjing, China, and also with the School of Informatics, University of Edinburgh, Edinburgh, UK (e-mail: wangyu1996@nuaa.edu.cn).}
\thanks{Zhaohui Zhu is with the College of Computer Science and Technology, Nanjing University of Aeronautics and Astronautics, Nanjing, China (e-mail: zhaohui@nuaa.edu.cn).}
\thanks{Rob van Glabbeek is with the School of Informatics, University of Edinburgh, Edinburgh, UK, and also with the School of Computer Science and Engineering, University of New South Wales, Sydney, Australia (e-mail: rvg@cs.stanford.edu). He is supported by Royal Society Wolfson Fellowship RSWF\textbackslash R1\textbackslash 221008.}
\thanks{Shigemasa Takai is with the Division of Electrical, Electronic and Infocommunications Engineering, Osaka University
(takai@eei.eng.osaka-u.ac.jp).}
\thanks{Jinjin Zhang is with the School of Computer Science, Nanjing Audit University, Nanjing, China (e-mail: jinjinzhang@nau.edu.cn).}
\thanks{Lixing Tan is with the College of Information Engineering, Taizhou University (e-mail: tanlixing@nuaa.edu.cn).}}

\maketitle

\vspace{-2ex}
\begin{abstract}
		
		Takai proposed a method for constructing a  maximally permissive supervisor for the similarity control problem (IEEE Transactions on Automatic Control,
		66(7):3197–3204, 2021).  This paper points out  that this construction does not (necessarily) work when the specification is not image-finite.  Inspired by Takai's construction, the notion of a (saturated)
		$(G,R)$-automaton is introduced and  theorems concerning (maximally permissive)
		supervisors for the similarity control problem are provided in terms of this notion. As an application of these theorems, the flaws in Takai's work are corrected.
\end{abstract}


\section{Introduction}

	Discrete event systems (DES) are event-driven systems composed of discrete events which happen at distinct points in time. Supervisory control theory, originating from the work of   Ramadge and Wonham \cite{1987Supervisory}, provides a framework for control of discrete event systems. In this theory \cite{1987Supervisory,Kimura2014Maximally,kushi2017synthesis,2019Maximally,2022Synthesis,2023MaximallyPermissive,SUN2014287,2020Synthesis,2018Nonblockingsimilarity,2007Control,2006Control,zhou2005small,9029606}, both plants and  specifications are  modeled as (non)deterministic automata. A supervisor controller, called supervisor, is also modeled as an automaton, which exerts control over a plant such that the supervised system runs as desired.
	
	In the development of supervisory control theory, various notions (e.g.\  language equivalence \cite{1987Supervisory}, bisimulation \cite{Kimura2014Maximally,SUN2014287,2006Control,zhou2005small} and simulation \cite{kushi2017synthesis,2019Maximally,2022Synthesis,2023MaximallyPermissive,2020Synthesis,2018Nonblockingsimilarity,2007Control,9029606}) have been adopted to formalize what it means that the behavior of a supervised system meets the requirements specified by a specification. Among them, the notion of simulation is suitable for handling situations with nondeterministic transitions and uncontrollable events.
	
	The similarity control problem asks, given a plant and specification, how to synthesize a supervisor for the former so that the supervised system is simulated by the latter  \cite{kushi2017synthesis,2022Synthesis,2023MaximallyPermissive,2020Synthesis,2006Control}. In \cite{kushi2017synthesis},  a necessary and sufficient condition for the existence of solutions to the similarity control problem is obtained. Ideally, the synthesized supervisor  should be as permissive as possible. To this end, Takai presents  		a method to synthesize a maximally permissive supervisor for the similarity control problem \cite{2020Synthesis}. Further, such a method is  extended to the (nonblocking) similarity control problem under partial event observations \cite{2019Maximally,9029606,2022Synthesis,2023MaximallyPermissive}. 
	
	The main contribution of our work is to  point out that  the method proposed in 
		\cite{2022Synthesis,2020Synthesis,9029606} for constructing a maximally permissive supervisor is not universally applicable when systems are not image-finite. 
	In detail, the results  in 
		\cite{2022Synthesis,2020Synthesis,9029606}  hold for automata with a finite number of states, but not for ones with infinitely many states. However, the latter are often used to model discrete event systems, such as in the study of supervisory control theory \cite{1987Supervisory}, protocols \cite{aarts2010generating} and counter automata \cite{valiant1975deterministic}. Hence, it is worth extending  the work in \cite{2022Synthesis,2020Synthesis,9029606} to the general case.
	Based on the method proposed in \cite{2020Synthesis}, we introduce the notion of a (saturated) $(G,R)$-automaton and show that any (saturated) $(G,R)$-automaton is a (maximally permissive) supervisor for the similarity control problem. Further, flaws in Takai's work are corrected.

	The rest of the paper is organized as follows. Some   preliminaries are recalled  in Section~\ref{sec:prel}.  Section~\ref{sec:counter} provides a counterexample for Theorem 1-3 in  \cite{2020Synthesis}. In Section~\ref{sec:GR}, the notion of a (saturated) $(G,R)$-automaton  and two theorems are proposed, and the flaws in \cite{2020Synthesis} are corrected.	Section~\ref{sec:flaws} briefly discusses the flaws occurring in~\cite{2022Synthesis,9029606}. We conclude this paper in Section~\ref{sec:conc}.

This paper is a slightly extended version of \cite{WZGTZT26}, which lacks Section~\ref{sec:flaws}.

\section{Preliminaries}\label{sec:prel}
	An  automaton  is  a 4-tuple $G=(X,\Sigma, \longrightarrow_G, X_0)$, where $X$ is the set of states, $\Sigma$ is the finite set of events, ${\longrightarrow_G}\subseteq X\times \Sigma\times X$ and $X_0\subseteq X$ is the nonempty set of initial states. 		
	As usual, we write $x\stackrel{\sigma}{\longrightarrow}_Gx'$ whenever $(x,\sigma, x')\in {\longrightarrow_G}$, $x\stackrel{\sigma}{\longrightarrow}_G$ iff $\exists\, x' (x\stackrel{\sigma}{\longrightarrow}_Gx')$, moreover, \plat{$x\,\,\,\,\not\!\!\!\!\stackrel{\sigma}{\longrightarrow}_G$} iff $\nexists\, x' (x\stackrel{\sigma}{\longrightarrow}_Gx')$.

	In the standard way, ${\longrightarrow_G}\subseteq X\times \Sigma\times X$ may be generalized to ${\longrightarrow_G}\subseteq X\times \Sigma^*\times X$, where $\Sigma^*$ is the set of all finite sequences of events in $\Sigma$.
	For any $s\in\Sigma^*$, a state $x$ is said to be  $s$-reachable  in an  automaton $G$ if $x_0\stackrel{s}{\longrightarrow_G} x$ for some $x_0\in X_0$.  
	Additionally, a state $x$ is said to be  reachable in $G$ if $x$ is  $s$-reachable   for some $s\in\Sigma^*$. In the following,   the subscript of the transitions (e.g.\  $G$ in $\longrightarrow_G$) will be omitted if this simplification does not cause any confusion. A state $x$ is a deadlock if, for each $\sigma\in \Sigma$, there is no $x'\in X$ such that $x\stackrel{\sigma}{\longrightarrow}x'$.
\pagebreak
	
		\textbf{Convention:}	In the remainder of this paper, a plant, supervisor and specification are modeled as automata $G=(X,\Sigma,\longrightarrow,X_0)$, $S=(Y,\Sigma,\longrightarrow, Y_0)$ and $R=(Z,\Sigma, \longrightarrow, Z_0)$ respectively. It is assumed by default that $G$, $R$ and $S$ have the aforementioned forms when they appear.

	\begin{myDef}\label{def-composition} \cite{2006Control}
		A plant $G$ supervised by a supervisor $S$ is modeled by the synchronous composition $S||G$ of $S$ and $G$, which is defined as $S||G=(Y\times X, \Sigma, \longrightarrow, Y_0\times X_0)$, where
		\begin{align*}
			&(y,x)\stackrel{\sigma}{\longrightarrow}(y_1,x_1) \text{ iff } x\stackrel{\sigma}{\longrightarrow}x_1 \text{ and }  y\stackrel{\sigma}{\longrightarrow}y_1 \\   		
			&\text{ for any } (y,x), (y_1,x_1)\in Y\times X \text{ and } \sigma\in \Sigma.
		\end{align*}
		
	\end{myDef}
	
	In  supervisory control theory, the events are usually divided into two classes: controllable  and uncontrollable events.  We denote by 	$\Sigma_{uc}$ (or, $\Sigma_{c}$) the set of uncontrollable (controllable, resp.) events, that is, $\Sigma=\Sigma_{uc}\cup\Sigma_{c}$ and $\Sigma_{uc}\cap\Sigma_{c}=\emptyset$.	
	In order to formalize the intuition that uncontrollable events are not disabled by  supervisors, the notion of $\Sigma_{uc}$-admissibility is adopted in \cite{{Kimura2014Maximally}}.

	\begin{myDef}\label{def-admissible l}
		Given a plant $G$, a supervisor $S$ is said to be $\Sigma_{uc}$-admissible with respect to $G$ if, for any reachable state $(y,x)$ in $S||G$, $(y,x)\stackrel{\sigma}{\longrightarrow}$ \text{whenever } $x\stackrel{\sigma}{\longrightarrow} \text{ and } \sigma\in \Sigma_{uc}.$
	\end{myDef}

    \vspace{-3ex}

	\begin{myDef}\label{def-conventional simulation}
		Let $G$ and $R$  be two  automata. A relation $\Phi\subseteq X\times Z$ is said to be a simulation (or, $\Sigma_{uc}$-simulation) relation from $G$ to $R$ if
		
		$\bullet$ for any $x_0\in X_0$, $(x_{0},z_{0})\in \Phi$ for some $z_0\in Z_0$,  and
		
		$\bullet$ for any $(x,z)\in \Phi$, $x'\in X$ and $\sigma\in \Sigma$ ($\sigma\in \Sigma_{uc}$, resp.), $x\stackrel{\sigma}{\longrightarrow}x'$ implies $z\stackrel{\sigma}{\longrightarrow}z'$ and $(x',z')\in \Phi$ for some $z'\in Z$.
	\end{myDef}
	
	If there exists a  simulation (or, $\Sigma_{uc}$-simulation) relation  from $G$ to $R$, then $G$ is said to be simulated ($\Sigma_{uc}$-simulated, resp.) by $R$, in symbols, $ G\sqsubseteq R$ ($G\sqsubseteq_{uc} R$, resp.). We write  $G\not\sqsubseteq R$ and $G\not\sqsubseteq_{uc} R$ for the negations of these statements.
 It is easy to see that both $\sqsubseteq$ and $\sqsubseteq_{uc}$ are  preorders, that is, they are reflexive and transitive.
	We also adopt the notation $\Phi:G\sqsubseteq R$ (or, $\Phi:G\sqsubseteq_{uc} R$) to indicate that $\Phi$ is a simulation ($\Sigma_{uc}$-simulation, resp.) relation from $G$ to $R$.
	
	Given a plant $G$ and specification $R$, the so called 
	\textit{$(G,R)$-similarity control problem} refers to  finding a $\Sigma_{uc}$-admissible (w.r.t.\  $G$) supervisor $S$ such that $S||G\sqsubseteq R$ \cite{kushi2017synthesis,2020Synthesis,2006Control}. This problem is said to be solvable if such a supervisor exists.
	For convenience, let $\SP(G,R)$ be the set of all $\Sigma_{uc}$-admissible (w.r.t.\  $G$) supervisors  which solve the $(G,R)$-similarity control problem, that is,   
$$
		\SP(G,R)= \{S: S||G \sqsubseteq R \text{ and } S \text{ is } \Sigma_{uc}\text{-admissible w.r.t.\  } G \}.
$$
	\noindent Moreover, let $\MSP(G,R)$ be the set of all maximally
	permissive supervisors for  the similarity control
	problem, that is,
	$$	\MSP(G, R)= \{S\mathbin\in \SP(G, R): \forall S'\mathbin\in \SP(G, R)(S'||G\sqsubseteq S||G)\}.$$

	In \cite{kushi2017synthesis},  a necessary and sufficient condition for the existence of a solution to the $(G,R)$-similarity control problem is given in terms of $\Sigma_{uc}$-similarity, namely, $\SP(G,R)\neq\emptyset$ iff $G\sqsubseteq_{uc}R$.
	Next we recall the method presented in \cite{2020Synthesis} for synthesizing a maximally permissive
	supervisor  that solves the similarity control problem.

	\begin{myDef} ($(G,R)^\uparrow$) \label{def-S arrow old}
		Given a plant $G$ and specification $R$, the function $F_{(G,R)}:\powerset(X\times Z)\footnote{Given a set $X$, $\powerset(X)$ denotes the set $\{X_1:X_1\subseteq X\}$.}\longrightarrow \powerset(X\times Z)$ is defined as, for each $W\subseteq X\times Z$,
		\begin{align*}
			& F_{(G,R)}(W)=\{(x,z)\in W: \forall \sigma\in \Sigma_{uc}\   \forall x'\in X\\ 			 
			&(x\stackrel{\sigma}{\longrightarrow}x' \Longrightarrow \exists z'\in Z(z\stackrel{\sigma}{\longrightarrow}z' \text{ and } (x',z')\in W ))\}.
		\end{align*}

		\noindent Based on the  greatest fixpoint of the function $F_{(G,R)}$ (denoted by $W_{(G,R)}^\uparrow$), the automaton $(G,R)^\uparrow=(Y^\uparrow, \Sigma, \longrightarrow,\linebreak Y_0^\uparrow)$ with $Y^\uparrow= \powerset(W_{(G,R)}^\uparrow)$ is defined as 
\begin{align*}
	Y_0^\uparrow=\{&W_0\in Y^\uparrow\cap \powerset(X_0\times Z_0): |W_0|=|X_0| \text{ and }\\[-1ex]
	&\forall x_0\in X_0\ \exists z_0\in Z_0((x_0,z_0)\in W_0)   \}\footnotemark.
\end{align*}
\footnotetext{Given a set $X$, $|X|$ denotes the cardinality of $X$.}
		\noindent and, for any $W$, $W'\in Y^\uparrow$ and $\sigma\in \Sigma$,  $W\stackrel{\sigma}{\longrightarrow} W'$ iff\pagebreak[2]
		
		$(\ref{def-S arrow old}-a)$ there exists $(x,z)\in W$ such that $x\stackrel{\sigma}{\longrightarrow}$;
		
		$(\ref{def-S arrow old}-b)$ either $\sigma\in \Sigma_{uc}$ or $	\forall(x,z)\in W\ \forall x'\in X ~ (x\stackrel{\sigma}{\longrightarrow} x'\linebreak
		\Longrightarrow \exists z'\in Z (z\stackrel{\sigma}{\longrightarrow}z'\text{ and } (x',z')\in  W_{(G,R)}^\uparrow))$;
		
		$(\ref{def-S arrow old}-c)$ $W'$ is minimal in $N(W,\sigma)$, that is, $W_1=W'$ for any $W_1\in N(W,\sigma)$ with $W_1\subseteq W'$, where $N(W,\sigma)=$\vspace{-2ex}
		\begin{align*}\nonumber
			\{&W''\in Y^\uparrow\cap\powerset\left(\bigcup_{(x,z)\in W}\{x':x\stackrel{\sigma}{\longrightarrow} x'\}\times\{z':z\stackrel{\sigma}{\longrightarrow} z'\}\right) :\\
			&\!\!\forall (x,z)\mathbin\in W \, \forall x'(x\mathbin{\stackrel{\sigma}{\longrightarrow}} x'\Rightarrow \exists z'(z\mathbin{\stackrel{\sigma}{\longrightarrow}}z'\text{ and }  (x',z')\mathbin\in W'') )
			\}.
		\end{align*}
	\end{myDef}

	Note that $N(W,\sigma)\neq\emptyset$ if $W$ and $\sigma$ satisfy clauses $(\ref{def-S arrow old}-a)$ and $(\ref{def-S arrow old}-b)$ in Definition \ref{def-S arrow old}.  Clause	$(\ref{def-S arrow old}-b)$	characterizes events that
		should be enabled at $W$.
	Each element of $N(W,\sigma)$ is a state in $Y^\uparrow$, and intuitively, it consists of candidate $\sigma$-labeled successors of $W$. As pointed out in \cite{2020Synthesis}, the motivation for adopting the minimal elements instead of all elements in $N(W,\sigma)$ as the $\sigma$-labeled successors in clause $(\ref{def-S arrow old}-c)$ is to reduce the number of possible destination states. 	It has been asserted that the initial state set $Y_0^\uparrow\neq \emptyset$ and $(G,R)^\uparrow\in \MSP(G,R)$ whenever $G\sqsubseteq_{uc}R$ (see Theorem 1, 2 and (2) in \cite{2020Synthesis}). Unfortunately, a counterexample given in the next section will show that this construction method based on minimal elements does not always work well. In fact, for automata with infinite state spaces, the automaton $(G,R)^\uparrow$ is not always $\Sigma_{uc}-$admissible even if $G\sqsubseteq_{uc}R$.

	\begin{proposition} \cite{Kimura2014Maximally}
		Let  $G$ and  $R$ be two automata.\label{prop-fixpoint}
		
		$(\ref{prop-fixpoint}-a)$ For any $\Phi\subseteq X\times Z$, $\Phi$ is a $\Sigma_{uc}$-simulation from $G$ to $R$ iff $\Phi=F_{(G,R)}(\Phi)$ and $\forall x_0\in X_0\  \exists z_0\in Z_0 ((x_0,z_0)\in \Phi)$.

		$(\ref{prop-fixpoint}-b)$ $G\sqsubseteq_{uc}R$ iff $\forall x_0\in X_0 \ \exists z_0\in Z_0 ((x_0,z_0)\in W_{(G,R)}^\uparrow)$.

		$(\ref{prop-fixpoint}-c)$ $W_{(G,R)}^\uparrow$ is a $\Sigma_{uc}$-simulation from $G$ to $R$ whenever $G\sqsubseteq_{uc}R$.
	\end{proposition}
	\vspace{-1ex}
	
	In the situation that $G\sqsubseteq_{uc}R$, clause $(\ref{def-S arrow old}-b)$ in Definition \ref{def-S arrow old} can be equivalently be simplified by omitting the disjunct $\sigma\in \Sigma_{uc}$. Formally,
	
	\vspace{-1ex}
	\begin{observation}\label{obs-1}
		Let $G$ and $R$ be two automata with $G\sqsubseteq_{uc}R$ and $(G,R)^\uparrow=(Y^\uparrow, \Sigma, \longrightarrow, Y_0^\uparrow)$. Then, for any $W\in Y^\uparrow$ and $\sigma\in\Sigma$, clause $(\ref{def-S arrow old}-b)$ in Definition \ref{def-S arrow old} is equivalent to
		\begin{align*}\label{equation-obs1}
			&\forall(x,z)\in W\ \forall x'\in X (x\stackrel{\sigma}{\longrightarrow} x'\Longrightarrow \\
			&\exists z'\in Z (z\stackrel{\sigma}{\longrightarrow}z'  \text{ and } (x',z')\in W_{(G,R)}^\uparrow) ).\tag{\ref{obs-1}-1} 		 
		\end{align*} 
        \vspace{-4ex}
		\begin{proof}
			($\Longleftarrow$) It holds trivially.
			
			($\Longrightarrow$)  Clearly, (\ref{equation-obs1}) holds whenever $\sigma\notin \Sigma_{uc}$. Next we deal with the case where  $\sigma\in \Sigma_{uc}$. For any  $(x,z)\in W\subseteq W_{(G,R)}^\uparrow$ and $x'\in X$ such that $x\stackrel{\sigma}{\longrightarrow} x'$, by clause $(\ref{prop-fixpoint}-c)$ in Proposition \ref{prop-fixpoint}  and Definition \ref{def-conventional simulation}, \plat{$z\stackrel{\sigma}{\longrightarrow}z'$} and $(x',z')\in W_{(G,R)}^\uparrow$ for some $z'\in Z$, as desired.
		\end{proof}
	\end{observation}

	To eliminate certain deadlocks   in the supervised plant 
	$(G,R)^\uparrow||G$, another candidate maximally permissive supervisor $\widetilde{(G,R)^\uparrow}$ for the similarity control problem is constructed as follows,  where a transition in $(G,R)^\uparrow$ like $W\stackrel{\sigma}{\longrightarrow}W'$ with $W'$ a deadlock is preserved only if all $\sigma$-reachable states from $W$ are deadlocks.

	\begin{myDef}\cite{2020Synthesis}\label{def-remove deadlocks}
		Let  $G$ be  a plant, $R$ a specification  and $(G,R)^\uparrow=(Y^\uparrow, \Sigma, \longrightarrow, Y_0^\uparrow)$. Then, the supervisor $\widetilde{(G,R)^\uparrow}$ is defined as $\widetilde{(G,R)^\uparrow}=(Y^\uparrow, \Sigma, \longrightarrow_{\sim}, Y_0^\uparrow)$, where, for any $W,W'\in Y^\uparrow$ and $\sigma\in \Sigma$, $W\stackrel{\sigma}{\longrightarrow}_{\sim}W'$ iff $W\stackrel{\sigma}{\longrightarrow}W'$ and
		
		$(\ref{def-remove deadlocks}-a)$ either	$W'$ is not a deadlock in $(G,R)^\uparrow$, or
		
		$(\ref{def-remove deadlocks}-b
		)$ $W_1$ is a deadlock in $(G,R)^\uparrow$ for any $W_1\in Y^\uparrow$ such that $W\stackrel{\sigma}{\longrightarrow}W_1$.
	\end{myDef}

	We recall a result obtained in \cite{kushi2017synthesis}, which will be used in Section~\ref{sec:GR}. 	It shows that  a simulation relation can be converted into a 
	$\Sigma_{uc}$-simulation through an appropriate projection.
	\vspace{-1ex}
	
	\begin{lemma}\label{lemma-couter-touying is  simulation}
		For any plant $G$ and  specification $R$, if $S$ is  a $\Sigma_{uc}$-admissible (w.r.t.\  $G$)  supervisor and $\Phi: S||G\sqsubseteq R$, then $\pi(\Phi)$ is a $\Sigma_{uc}$-simulation from $G$ to $R$, where 
		\vspace{-0.5cm}

		\begin{align*}\label{equation-pi}
			\pi(\Phi)=&\{(x,z): ((y,x),z)\in \Phi \text{ and }\\[-2pt]
			& (y,x) \text{ is  reachable in }S||G  \text{ for some } y\in Y\}.	\notag
		\end{align*}		
	\end{lemma}      	
	
	We end this section by providing a discrete event system which can only be modeled using automata with infinitely many states.

	\begin{example}\label{example-automata infinite state}
		 The system $G$ randomly outputs nonempty palindromes in $\{0,1\}^*$ with an even length. Here, a string		$w\in\{0,1\}^*$ is said to be a palindrome if it is the same as its reversal, i.e.,  $w=w^R$, where $w^R=a_na_{n-1}...a_1a_0$ for $w=a_0a_1...a_{n-1}a_n$. Clearly, all strings that can be outputted by $G$ have the form like $ww^R$ with $(\emptyset\neq) w\in\{0,1\}^*$. It is easy to see that $G$ can be specified by the automaton with infinitely many states over the event set $\Sigma=\{0,1\}$ displayed in Figure \ref{figure-application specification1}, where $0$ and $1$ stand for the events `$output$ $0$' and `$output$ $1$' respectively. By the well-known Pumping Lemma in the theory of automata and formal languages (see, e.g., \cite{10.1145/568438.568455}), the language $\{ww^R:w\in\{0,1\}^* \text{ and } w\neq \emptyset\}$ is not regular, and hence it can not be accepted by any finite automata.  Therefore, the system $G$ can not be modeled as a finite automaton.

		\begin{figure}
			\centering
			
			\begin{minipage}[t]{0.34\textwidth}
				\vspace{0pt}
				\begin{tikzpicture}[node distance=1cm]
					\tikzstyle{circleNode} = [circle, draw, inner sep=0pt, minimum size=4pt]
					\tikzstyle{arrow} = [->, >=stealth, shorten >=1pt]		
					\node (x0) [circleNode] {$x_{0}$};

					\node (x13) [circleNode, right of=x0, yshift=-1.2cm] {$x_{13}$};
					\node (x23) [circleNode, right of=x13] {$x_{23}$};
					\node (x33) [circleNode, right of=x23] {$x_{33}$};
					\node (x43) [circleNode, right of=x33] {$x_{43}$};
					\node (x53) [circleNode, right of=x43] {$x_{53}$};
					\node (x63) [circleNode, right of=x53] {$x_{63}$};
					
					% 中间分支保持不变
					\node (x12) [circleNode, right of=x0] {$x_{12}$};
					\node (x22) [circleNode, right of=x12] {$x_{22}$};
					\node (x32) [circleNode, right of=x22] {$x_{32}$};
					\node (x42) [circleNode, right of=x32] {$x_{42}$};

					\node (x11) [circleNode, right of=x0, yshift=1.2cm] {$x_{11}$};
					\node (x21) [circleNode, right of=x11] {$x_{21}$};

					% 调整箭头
					% 对于下侧分支（现 x13 分支），箭头标签放在下侧
					\draw [arrow] (x0) -- (x13) node[midway, below]{$0$};
					\draw [arrow] (x13) -- (x23) node[midway, below]{$1$};    
					\draw [arrow] (x23) -- (x33) node[midway, below]{$1$};
					\draw [arrow] (x33) -- (x43) node[midway, below]{$1$};            
					\draw [arrow] (x43) -- (x53) node[midway, below]{$1$};            
					\draw [arrow] (x53) -- (x63) node[midway, below]{$0$};
					
					% 中间分支箭头保持不变
					\draw [arrow] (x0) -- (x12) node[midway, above]{$0$};
					\draw [arrow] (x12) -- (x22) node[midway, above]{$1$};
					\draw [arrow] (x22) -- (x32) node[midway, above]{$1$};
					\draw [arrow] (x32) -- (x42) node[midway, above]{$0$};
					
					% 对于上侧分支（现 x11 分支），箭头标签放在上侧
					\draw [arrow] (x0) -- (x11) node[midway, above]{$0$};
					\draw [arrow] (x11) -- (x21) node[midway, above]{$0$};         
					\node (cdots1) [below of=x0, xshift=0.2cm] {$\vdots$};
					
				\end{tikzpicture}
			\end{minipage}
			\vspace{2ex}	
			\caption{The system $G$} 
			\label{figure-application specification1} 
		\end{figure}
	\end{example}

	Let's  review the well-known concept below (see, e.g.,~\cite{1980Robin}), which characterizes automata with only finitely many 
		$\sigma$-labeled successors in any state for each $\sigma$.

	\begin{myDef}(Image-finite)\label{def-image-finite}
		An automaton $G$ is said to be image-finite if the set $\{x':x\stackrel{\sigma}{\longrightarrow}x'\}$ is finite for any $x\in X$ and $\sigma\in \Sigma$.
	\end{myDef}

The automaton in Figure \ref{figure-application specification1} is not image-finite. However, through introducing infinitely many initial states, $G$ can be modeled by image-finite automata. For example, $G$ can be modeled as shown in Figure \ref{figure-application specification2}, where {$x_{0i}$ ($i\geqslant 0$)} are initial states.
	\begin{figure}[ht]
		\begin{minipage}[t]{\linewidth}		
			\centering	
			\vspace{0pt}
			\begin{tikzpicture}[node distance=1cm]
				\tikzstyle{circleNode} = [circle, draw, inner sep=0pt, minimum size=4pt]
				\tikzstyle{arrow} = [->, >=stealth, shorten >=1pt]		
				
				% First path (x01 -> x11 -> x21)
				\node (x01) [circleNode] {$x_{01}$};
				\node (x11) [circleNode, right of=x01] {$x_{11}$};					
				\node (x21) [circleNode, right of=x11] {$x_{21}$};
				\draw [arrow] (x01) -- (x11) node[midway, below]{\scriptsize $0$};
				\draw [arrow] (x11) -- (x21) node[midway, below]{\scriptsize $0$};
				
				% Second path (x02 -> x12 -> x22 -> x32 -> x42)
				\node (x02) [circleNode, below of=x01] {$x_{02}$};
				\node (x12) [circleNode, right of=x02] {$x_{12}$};					
				\node (x22) [circleNode, right of=x12] {$x_{22}$};
				\node (x32) [circleNode, right of=x22] {$x_{32}$};
				\node (x42) [circleNode, right of=x32] {$x_{42}$};
				\draw [arrow] (x02) -- (x12) node[midway, below]{\scriptsize $0$};
				\draw [arrow] (x12) -- (x22) node[midway, below]{\scriptsize $1$};
				\draw [arrow] (x22) -- (x32) node[midway, below]{\scriptsize $1$};
				\draw [arrow] (x32) -- (x42) node[midway, below]{\scriptsize $0$};
				
				% Third path (x03 -> x13 -> x23 -> x33 -> x43 -> x53 -> x63)
				\node (x03) [circleNode, below of=x02] {$x_{03}$};
				\node (x13) [circleNode, right of=x03] {$x_{13}$};					
				\node (x23) [circleNode, right of=x13] {$x_{23}$};
				\node (x33) [circleNode, right of=x23] {$x_{33}$};
				\node (x43) [circleNode, right of=x33] {$x_{43}$};
				\node (x53) [circleNode, right of=x43] {$x_{53}$};
				\node (x63) [circleNode, right of=x53] {$x_{63}$};
				\draw [arrow] (x03) -- (x13) node[midway, below]{\scriptsize $0$};
				\draw [arrow] (x13) -- (x23) node[midway, below]{\scriptsize $1$};
				\draw [arrow] (x23) -- (x33) node[midway, below]{\scriptsize $1$};
				\draw [arrow] (x33) -- (x43) node[midway, below]{\scriptsize $1$};	
				\draw [arrow] (x43) -- (x53) node[midway, below]{\scriptsize $1$};				
				\draw [arrow] (x53) -- (x63) node[midway, below]{\scriptsize $0$};	
				
				% Add dots for continuity
				\node (cdots1) [below of=x03, yshift=0.4cm] {$\vdots$};
				
			\end{tikzpicture}
		\end{minipage}
		\caption{Another model for $G$} 
		\label{figure-application specification2} 
	\end{figure}
In our view, this is far less natural than the automaton shown in Figure \ref{figure-application specification1}. Introducing infinitely many initial states is the price that must be paid to obtain image-finiteness. By a trivial application of K\"{o}nig's Lemma (see, e.g., \cite{10.5555/2222837}), there is no image-finite automaton $G'$ with finitely many initial states such that the completed trace set\footnote{Here a \emph{completed trace} of an automaton \cite{vG01} is any sequence of labels of an execution path that either is infinite or ends in a state from which no further transitions are possible.} of $G'$ is $\{ww^R:w\in\{0,1\}^* \text{ and } w\neq \emptyset\}$.

\section{A Counterexample}\label{sec:counter}
	
Theorem 1-3 in  \cite{2020Synthesis} assert that, for any plant $G$ and specification $R$, if $G\sqsubseteq_{uc} R$, then  $(G,R)^\uparrow, \widetilde{(G,R)^\uparrow} \in \SP(G, R)$ and  $(G,R)^\uparrow,\widetilde{(G,R)^\uparrow}\in \MSP(G, R)$, that is, both  $(G,R)^\uparrow$ and $\widetilde{(G,R)^\uparrow}$ are maximally permissive supervisors for  the $(G,R)$-similarity control problem. Unfortunately, none of these three theorems is valid in general. This section intends to show this 
by providing a counterexample below.

\begin{example}\label{example-couter example}
	Consider the   specification $R$ and plant $G$  given in Figure \ref{figure-counterexample1}, where all the transitions  are labeled by the event $\sigma\in \Sigma_{uc}$. 
	
	\begin{figure}[htbp]\vspace{-4ex}
		\centering
		\includegraphics[scale=0.6]{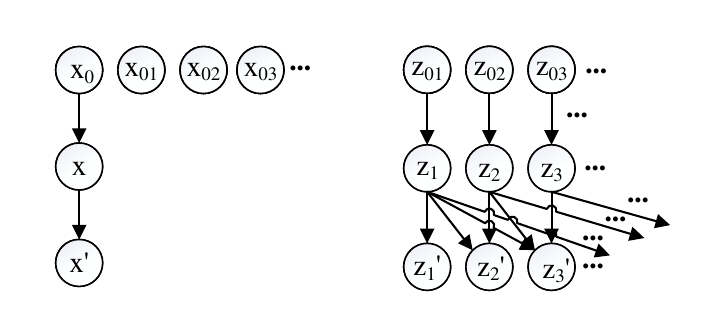}
		\vspace{-0.1cm}   	
		\caption{The plant $G$ (left) and specification $R$ (right)}
		\label{figure-counterexample1}
		\vspace{-1pt}
	\end{figure}
	
	For the plant $G$, $X_0=\{x_0\}\cup \{x_{0i}:i\geqslant 1\}$ and $X=X_0\cup \{x,x'\}$, although      	
	there are infinitely many initial states $x_0, x_{01},x_{02},x_{03},...$;  each $x_{0i}$ $(i\geqslant1)$ is a deadlock and only $x_0$ enables a $\sigma$-labeled transition to the state $x$ and then to $x'$. 
	For the specification $R$, there are also infinitely many initial states $z_{01}, z_{02},z_{03},...$, all of which could perform one $\sigma$-labeled transition. Moreover,  each $z_i$ $(i\geqslant1)$ enables   $\sigma$-labeled transitions to $z_{k}'$ for all $k\geqslant i$.

	It can be checked straightforwardly that $G\sqsubseteq_{uc} R$, and thus, by Theorem 1 in  \cite{kushi2017synthesis}, there exists a supervisor $S$ such that $S||G\sqsubseteq R$.   For example, such a supervisor $S$ is given in Figure \ref{figure-counterexample4}. Moreover,  the reachable part of the supervised plant $S||G$  is also represented graphically in this figure, where all the transitions  are also labeled by the event $\sigma\in \Sigma_{uc}$. 
    
    \begin{figure}[htbp]\centering
		\includegraphics[scale=0.5]{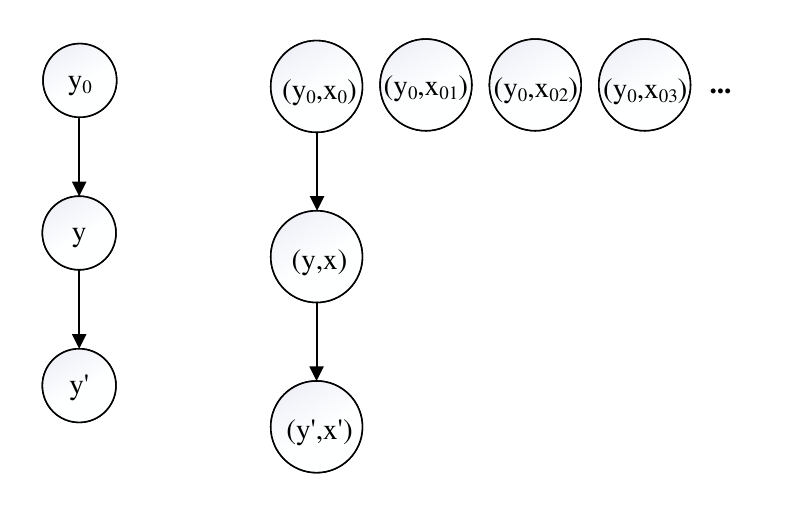}
		\caption{The supervisor $S$ (left)  and  reachable part of the supervised plant  $S||G$ (right)}
		\label{figure-counterexample4}
	\end{figure} 
    
    The greatest fixpoint $W_{(G,R)}^\uparrow$ of the function $F_{(G,R)}$  is
	$X\times Z-((\{x_0\}\times    \{z_i,z_i':i\mathbin{\geqslant} 1       \})\cup (\{x\}\times  \{z_i':i\mathbin{\geqslant} 1       \}))$.\linebreak[3] 
	In the following, we intend to demonstrate that neither $(G,R)^\uparrow$ nor $\widetilde{(G,R)^\uparrow}$ is $\Sigma_{uc}$-admissible, so $(G,R)^\uparrow\notin MSP(G,R)$ and $\widetilde{(G,R)^\uparrow}\notin MSP(G,R)$, which shows that none of Theorems 1-3 in  \cite{2020Synthesis} is valid.	
	
	By Definition \ref{def-S arrow old}, there are infinitely many initial states in $(G,R)^\uparrow$. For the state $W_0=\{(x_{0i},z_{0i})\!:i\mathbin\geqslant 1\}\cup \{(x_0,z_{0i})\!:i\mathbin\geqslant 1\}\linebreak[3]\subseteq W_{(G,R)}^\uparrow,$ it is obvious that $|W_0|=|X_0|$ and, for any $x\in X_0$, there exists $z\mathbin\in Z_0 \text{ such that } (x,z)\mathbin\in W_0$.    Hence, \plat{$W_0\mathbin\in Y_0^\uparrow$}.
  Moreover, $N(W_0,\sigma)\mathbin=\{W\}$ and $W_0\mathbin{\stackrel{\sigma}{\longrightarrow}}W$,
for the set $W=\{(x,z_i):i\geqslant 1\}\in Y^\uparrow$. 
	
	Clearly, $N(W,\sigma)$ is the set of all infinite subsets of $\{(x',z_i'):i\geqslant1\}$, and hence there exists no minimal element in it. Nevertheless, by clause $(\ref{def-S arrow old}-c)$, each $\sigma$-labeled successor of $W$ in $(G,R)^\uparrow$ must be a minimal element of $N(W,\sigma)$.
	Thus, $W$ is a deadlock     in $(G,R)^\uparrow$, and hence $(W,x)$ is also a deadlock   in  $(G,R)^\uparrow||G$.     
	Note that
	$(W_0,x_0)\stackrel{\sigma}{\longrightarrow}(W,x)$ due to $W_0\stackrel{\sigma}{\longrightarrow}W$ and $x_0\stackrel{\sigma}{\longrightarrow}x$, which means that $(W,x)$ is  reachable in $(G,R)^\uparrow||G$. However, since $x\stackrel{\sigma}{\longrightarrow}x'$, $\sigma\in \Sigma_{uc}$ and $(W,x)$ is a deadlock, by Definition \ref{def-admissible l}, $(G,R)^\uparrow$ is not  $\Sigma_{uc}$-admissible w.r.t.\  $G$, which implies $(G,R)^\uparrow\notin \SP(G, R)$, and thus $(G,R)^\uparrow\notin \MSP(G, R)$.
	
	Moreover, since $W$ is the unique successor of $W_0$ and $W$ is a deadlock  in $(G,R)^\uparrow$, by   Definition \ref{def-remove deadlocks}, $W_0\stackrel{\sigma}{\longrightarrow}_{\sim}W$ holds in $\widetilde{(G,R)^\uparrow}$. Obviously, $W$ is also a deadlock in $\widetilde{(G,R)^\uparrow}$.\vspace{-1pt} Similarly,\vspace{-1pt} it can be verified  that $\widetilde{(G,R)^\uparrow}$ is not  $\Sigma_{uc}$-admissible w.r.t.\  $G$, so $\widetilde{(G,R)^\uparrow}\mathbin{\notin} \SP(G, R)$ and $\widetilde{(G,R)^\uparrow}\mathbin{\notin} \MSP(G, R)$.
\end{example}

\vspace{-3ex}

\section[Saturated (G,R)-automata]{Saturated $(G,R)$-automata}\label{sec:GR}

In order to correct  Theorem 1-3 in \cite{2020Synthesis}, this  section will provide two theorems on (maximally permissive) supervisors.
Although Example \ref{example-couter example} illustrates that Takai's construction is not universally applicable, such a construction inspires the definition below, which  introduces a class of automata as  candidate maximally permissive supervisors for the similarity control problem.

\begin{myDef}\label{def-S arrow new}
	Given a plant $G$ and specification $R$,  an  automaton $\mathcal{A}=(A, \Sigma, \longrightarrow, A_0)$ with $A_0\neq \emptyset$ is said to be a $(G,R)$-automaton if it satisfies the following:
	
	(state)  $A=\powerset(W_{(G,R)}^\uparrow)$;
	
	(istate)  $A_0$ $\subseteq$ $\{W\in A\cap\powerset(X_0\times Z_0):\forall x_0\in X_0\  \exists z_0\in Z_0\linebreak((x_0,z_0)\in W)\}$;

	$(\ref{def-S arrow new}-a)$ for any reachable state $W\in A$ and $\sigma\in \Sigma$, $W\stackrel{\sigma}{\longrightarrow}$ whenever $W$ and $\sigma$ satisfy clauses $(\ref{def-S arrow old}-a)$ and $(\ref{def-S arrow old}-b)$ in Definition \ref{def-S arrow old};
	
	$(\ref{def-S arrow new}-b)$ for any reachable state $W$, $W'\in A$ and $\sigma\in \Sigma$, $W\stackrel{\sigma}{\longrightarrow} W'$ implies $W'\in N(W,\sigma)$.  
	
	\noindent	 Further,  $\mathcal{A}$ is said to be saturated  if 
	
	(sistate)  $\{W\in A\cap\powerset(X_0\times Z_0):\forall x_0\in X_0\  \exists !\footnote{The meaning of the quantifier $\exists !$ is ``there exists a unique object''.}z_0\in Z_0\linebreak((x_0,z_0)\in W)\}\subseteq$  $A_0$;
	
	$(\ref{def-S arrow new}-c)$ for any $W\in A$, $\sigma\in \Sigma$ and $W'\in N(W,\sigma)$, if $W$ is reachable in $\mathcal{A}$ and $W\stackrel{\sigma}{\longrightarrow}$, then, there exists $W''\in N(W,\sigma)$ such that $W\stackrel{\sigma}{\longrightarrow} W''$ and $W''\subseteq W'$.
\end{myDef}

The clauses $(\ref{def-S arrow new}-a,b,c)$ propose requirements for transitions. The condition $(\ref{def-S arrow new}-a)$ ensures that $W$ has $\sigma$-successors in certain scenarios, $(\ref{def-S arrow new}-b)$ limits the scope of $\sigma$-successors and $(\ref{def-S arrow new}-c)$ requires the existence of sufficient successors to ensure that the saturated automaton can simulate the behaviors of all potential solutions.
From the counterexample in Section\ref{sec:counter}, it can be observed that when the specifications and plants are modeled as automata with infinitely many states, the sets $N(W,\sigma)$ may also be infinite, so that it is not guaranteed that there 
	exist minimal elements in it.
To make up for this deficiency, condition $(\ref{def-S arrow new}-c)$ relaxes the requirements for minimal attributes of $\sigma$-successor states in clause 
	$(\ref{def-S arrow old}-c)$. It is not difficult to see that, for $N(W,\sigma)$ without infinite decreasing chains, 
	$(\ref{def-S arrow old}-c)$ provides a way to realize 
	$(\ref{def-S arrow new}-c)$.	

The reader will find that the notion above enables us to capture the structural requirements that supervisors and maximally permissive ones need to meet, respectively.
Intuitively, the saturation conditions (sistate) and $(\ref{def-S arrow new}-c)$ require a saturated $(G,R)$-automaton to contain sufficient initial states and transitions, which stems from our motivation that saturated automata will be treated as a class of maximally permissive supervisors. 
By the way, the automaton $(G,R)^\uparrow$ is not always a $(G,R)$-automaton. For instance, the one given in Example~\ref{example-couter example} doesn't satisfy clause $(\ref{def-S arrow new}-a)$ in Definition \ref{def-S arrow new}, because the reachable state $W$ contains the pair $(x,z_1)$ with \plat{$x\stackrel{\sigma}{\longrightarrow}$} and $\sigma\in \Sigma_{uc}$ but $W\,\,\,\,\not\!\!\!\!\stackrel{\sigma}{\longrightarrow}$.

Now we analyze the computational complexity of determining whether an automaton is a saturated $(G,R)$-automaton. Note that $G$ and $R$ are implicitly assumed to be finite automata in this context.
Given an automaton $\mathcal{B}=(B,\Sigma,\longrightarrow,B_0)$, the most expensive step of verifying whether it's a $(G,R)$-automaton is $(\ref{def-S arrow new}-b)$. Assume $\mathcal{B}$ satisfies the condition (state).
Given $W \in B$ and $\sigma\in\Sigma$, the time complexity of checking membership of the set $N(W,\sigma)$ is $O(|X|^2\times |Z|^2)$. There are at most $2^{2(|X|\times |Z|)}\!\times |\Sigma|$ transitions $W \stackrel{\sigma}{\longrightarrow} W'$, and for each of them this check has to be performed, yielding a total time cost of  $O( 2^{2(|X|\times |Z|)}\times |\Sigma| \times |X|^2\times |Z|^2)$.
Further, for a $(G,R)$-automaton, the most expensive step of checking whether it's saturated is the one of verifying $(\ref{def-S arrow new}-c)$.
Given $\sigma$, for each appropriate pair $W,W'\in B$ for which we find that $W' \in N(W,\sigma)$, we have to find an appropriate $W''\in B$ in the above-mentioned list of transitions. Hence, the complexity of verifying $(\ref{def-S arrow new}-c)$ is $O( 2^{3|X|\times |Z|)} \times|\Sigma| \times |X|^2\times |Z|^2).$ Overall, the  computational complexity of verifying whether $\mathcal{B}$ is a saturated $(G,R)$-automaton is $O( 2^{3(|X|\times |Z|)} \times|\Sigma| \times |X|^2\times |Z|^2)$.

\begin{example}\label{example-class example}
	Consider the plant $G$ and specification $R$   shown in Figure \ref{class example1}, 
	\begin{figure}[htbp]\centering
		\includegraphics[scale=0.6]{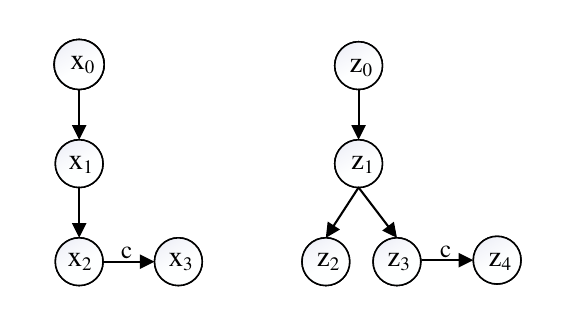}
		\caption{The plant $G$ (left) and specification $R$ (right)}
		\label{class example1}
	\end{figure}
	\noindent	where $c\mathbin\in \Sigma_{c}$ and all the omitted labels are $\sigma\mathbin\in \Sigma_{uc}$. Obviously, $G\sqsubseteq_{uc}R$ and 
		$W_{(G,R)}^\uparrow=\{(x_0,z_0),(x_1,z_1), (x_1,z_0)\}$
	 	$\cup
	 	\{(x_j,z_i):2\leqslant j\leqslant3 \text{ and } 0\leqslant i\leqslant 4\}.$

	The reachable parts of  the $(G,R)$-automata $\mathcal{A}$, $\mathcal{B}$ and $\mathcal{C}$ are shown in Figure \ref{class example2}, where $W_0=\{(x_0,z_0)\}$, $W_1=\{(x_1,z_1)\}$, $W_2=\{(x_2,z_2)\}$, $W_3=\{(x_2,z_3)\}$, $W_4=\{(x_2,z_2), (x_2,z_3)\}$ and $W_5=\{(x_3,z_4)\}$.
\begin{figure}[t]\vspace{-2ex}\centering
		\includegraphics[scale=0.6]{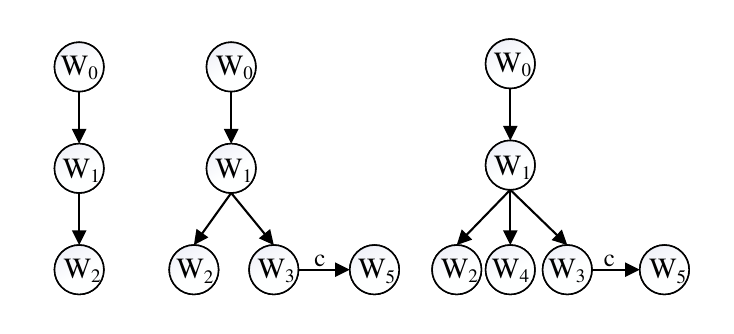}
		\caption{The reachable parts of  $(G,R)$-automata $\mathcal{A}$ (left) and saturated $(G,R)$-automata  $\mathcal{B}$ (middle) and $\mathcal{C}$ (right)}
		\label{class example2}
	\end{figure}
Due to $N(W_0,\sigma)=\{W_1\}$, $N(W_1,\sigma)=\{W_2,W_3,W_4\}$ and $N(W_3,c)=\{W_5\}$, the automaton $\mathcal{A}$ is  not saturated because clause $(\ref{def-S arrow new}-c)$ in Definition \ref{def-S arrow new} doesn't hold. For example, consider the reachable state $W_1$
		with $W_1\stackrel{\sigma}{\longrightarrow}$ in $\mathcal{A}$. Since $W_3\in N(W_1,\sigma)$ and there exists no $W^*\in N(W_1,\sigma)$ such that $W_1\stackrel{\sigma}{\longrightarrow}W^*$ with $W^*\subseteq W_3$, clause $(\ref{def-S arrow new}-c)$ doesn't hold for $W_1$, $\sigma$ and $W_3$. One can verify that $\mathcal{B}$ and $\mathcal{C}$ are saturated.
\end{example}

In the following, we intend to show that each (saturated) $(G,R)$-automaton is a (maximally permissive) supervisor.     It should be pointed out that our proof methods are from \cite{2020Synthesis}. We begin with  giving the following lemma,  which is  proposed in \cite{2020Synthesis} and holds for the automaton $(G,R)^\uparrow$. Since it only depends on the conditions (istate) and $(\ref{def-S arrow new}-b)$ in Definition \ref{def-S arrow new}, it is  still valid for any  $(G,R)$-automaton. For the sake of completeness, we provide its proof given in \cite{2020Synthesis}.

\vspace{-1ex}

\begin{lemma} \label{lemma-couter-pi(X)w2}
	Given a plant $G$ and  specification $R$, if $\mathcal{A}=(A,\Sigma,\longrightarrow,A_0)$ is a $(G,R)$-automaton, then, for  any $s\in \Sigma^*$, $W_0\in A_0$ and $W\in A$,  $W_0\stackrel{s}{\longrightarrow}W$ implies $\pi_G(W)=Reach(s,X_0)$, where
	\begin{align}
		&\pi_G(W)= \{x\in X: (x,z)\in W \text{ for some }z\in Z\} \text{ and }\notag\\ &Reach(s,X_0)=\{x\in X: x_0\stackrel{s}{\longrightarrow}x\text{ for some }x_0\in X_0\}\notag.
	\end{align}

	\begin{proof}
		It proceeds by induction on $|s|$, i.e., the length of $s$. If $|s|=0$ then $s=\varepsilon$, and hence $\pi_G(W_0)=X_0=Reach(s,X_0)$ by clause (istate) in Definition \ref{def-S arrow new}. In the following, we deal with the inductive step and let $|s|=k+1$.   		 		
		Assume that $s=s_1\sigma$ and $W_0\stackrel{s_1}{\longrightarrow}W_1\stackrel{\sigma}{\longrightarrow}W$ with  $W_1\in A$. By the induction hypothesis, we have $\pi_G(W_1)=Reach(s_1,X_0)$. Next we show $\pi_G(W)=Reach(s,X_0)$. 
		
		($\subseteq$) Let $x\in \it left.$ Then, there exists $z\in Z$ such that $(x,z)\in W$. Due to $W_1\stackrel{\sigma}{\longrightarrow}W$, by clause $(\ref{def-S arrow new}-b)$ in Definition \ref{def-S arrow new}, we get
		$W\in N(W_1,\sigma)$, and hence	$$W\in\powerset(\bigcup_{(x_1,z_1)\in W_1}\{x':x_1\stackrel{\sigma}{\longrightarrow} x'\}\times\{z':z_1\stackrel{\sigma}{\longrightarrow} z'\}).$$
		Thus, 	there exists $(x_1,z_1)\in W_1$ such that $x_1\stackrel{\sigma}{\longrightarrow}x$ and $z_1\stackrel{\sigma}{\longrightarrow}z$. Due to $(x_1,z_1)\in W_1$, $x_1\in \pi_G(W_1)=Reach(s_1,X_0)$. Then, it follows that $x_0\stackrel{s_1}{\longrightarrow}x_1\stackrel{\sigma}{\longrightarrow}x$ for some $x_0\in X_0$, which implies $x\in Reach(s,X_0)$.
		
		$(\supseteq)$ Let $x\in right.$ Thus, $x_0\stackrel{s_1}{\longrightarrow}x_1\stackrel{\sigma}{\longrightarrow}x$ for some $x_1\in X$ and $x_0\in X_0$. Hence, $x_1\in Reach(s_1,X_0)=\pi_G(W_1)$, and thus $(x_1,z_1)\in W_1$ for some $z_1\in Z$. Due to $W_1\stackrel{\sigma}{\longrightarrow}W$, by clause $(\ref{def-S arrow new}-b)$ in Definition \ref{def-S arrow new}, $W\in N(W_1,\sigma)$. Further, it follows from  $x_1\stackrel{\sigma}{\longrightarrow}x$ and $(x_1,z_1)\in W_1$ that $z_1\stackrel{\sigma}{\longrightarrow}z$ and $(x,z)\in W$ for some $z\in Z$, and thus $x\in \pi_G(W)$.	
	\end{proof}

\end{lemma}

Next we show that   any $(G,R)$-automaton is a solution of the $(G,R)$-similarity control problem whenever $G\sqsubseteq_{uc} R$.

\begin{theorem}\label{th-couter-S arrow is a solution2}
	Given two automata $G$ and  $R$  with $G \sqsubseteq_{uc} R$, let $\mathcal{A}=(A,\Sigma,\longrightarrow,A_0)$ be any $(G,R)$-automaton. Then  $\mathcal{A}\in \SP(G,R) $.

	\begin{proof}
		Let  $\Phi_{uc}:G \sqsubseteq_{uc} R$.
		First we show the $\Sigma_{uc}$-admissibility of $\mathcal{A}$.    Let $(W,x)$ be a reachable state in $\mathcal{A}||G$ and $x\stackrel{\sigma}{\longrightarrow}$ with $\sigma\in \Sigma_{uc}$. 
		By Definition \ref{def-composition} and \ref{def-admissible l}, it suffices to show $W\stackrel{\sigma}{\longrightarrow}$.			
		Since $(W,x)$ is  reachable, there exists $(W_0,x_0)\in A_0\times X_0$ such that $(W_0,x_0)\stackrel{s}{\longrightarrow}(W,x)$ for some $s\in \Sigma^*$. Thus, $W_0\stackrel{s}{\longrightarrow}W$ and $x_0\stackrel{s}{\longrightarrow}x$ due to Definition \ref{def-composition}. Then, it follows from Lemma \ref{lemma-couter-pi(X)w2} that $x\in Reach(s,X_0)=\pi_G(W)$, and hence $(x,z)\in W$ for some $z\in Z$. Further, by $x\stackrel{\sigma}{\longrightarrow}$  and $\sigma\in \Sigma_{uc}$, it is easy to see that $W$ and $\sigma$ satisfy clauses $(\ref{def-S arrow old}-a)$ and $(\ref{def-S arrow old}-b)$ in Definition \ref{def-S arrow old}. Since $\mathcal{A}$ is an $(G,R)$-automaton, by clause $(\ref{def-S arrow new}-a)$ in Definition \ref{def-S arrow new}, we have $W\stackrel{\sigma}{\longrightarrow}$, as desired.

		To complete the proof, it remains to be proven that   $\mathcal{A}||G \sqsubseteq R$.  We define the relation $\Phi\subseteq(A\times X)\times Z $ as follows:
		\begin{align*}
			\Phi\mathbin= \{	&((W,x),z)\!: (x,z)\in W \text{ and } (W,x) \text{ is reachable in }    \mathcal{A}||G     \}.  
		\end{align*}   		 		
		Next we  show that $\Phi$ is a simulation relation from $ \mathcal{A}||G$ to $R$.  Let $(W_0,x_0)\in A_0\times X_0$. Then, $W_0\in A_0$ and $x_0\in X_0$. By clause (istate) in Definition \ref{def-S arrow new}, we have $(x_0,z_0)\in W_0$ for some $z_0\in Z_0$.  Thus, $((W_0,x_0),z_0)\in \Phi$. Hence, each initial state in $\mathcal{A}||G$ is related to one in $R$ via $\Phi$.
		
		Let $((W,x),z)\in\Phi$ and $(W,x)\stackrel{\sigma}{\longrightarrow}(W',x')$ with $\sigma\in \Sigma$. Then,  $(x,z)\in W$ and   $(W,x)$  is reachable in $\mathcal{A}||G$. Since $\mathcal {A}$ is a $(G,R)$-automaton and $W\stackrel{\sigma}{\longrightarrow}W'$, by clause $(\ref{def-S arrow new}-b
		)$   in Definition \ref{def-S arrow new}, we get $W'\in N(W,\sigma)$. Further, it follows from  $x\stackrel{\sigma}{\longrightarrow}x'$ and $(x,z)\in W$ that $z\stackrel{\sigma}{\longrightarrow}z'$ and $(x',z')\in W'$  for some $z'\in Z$. Clearly, $(W',x')$ is reachable in $\mathcal{A}||G$, and hence $((W',x'),z')\in\Phi$, as desired.
	\end{proof}
\end{theorem}

In the following, we show that each saturated $(G,R)$-automaton is  a maximally permissive supervisor for the $(G,R)$-similarity control problem.

\begin{theorem}\label{theorem-couter-maximall-permessive} 	Let $G$ and $R$ be two automata and $\mathcal{A}=(A, \Sigma, \longrightarrow, A_0)$ a saturated $(G,R)$-automaton. If $G\sqsubseteq_{uc} R$ then $S||G\sqsubseteq \mathcal{A}||G$	 for any $S\in \SP(G,R)$.

	\begin{proof}    		
		Let $S\in \SP(G,R)$ and $\Phi: S||G\sqsubseteq R$. Since $\mathcal{A}$ is a  $(G,R)$-automaton, by Theorem \ref{th-couter-S arrow is a solution2}, we have $\mathcal{A}\in \SP(G,R)$. To complete the proof, it suffices to show that the relation $\Psi\subseteq (Y\times X)\times (A\times X)$ defined as				
		\begin{align*}\nonumber
			\Psi=\{&((y,x),(W,x)):\  \forall(x_W,z_W)\in W (((y,x_W),z_W)\in \Phi)\\  
			&\text{ and }
			\exists s\in \Sigma^*  				
			((y,x) \text{ and } (W,x)  \text{ are $s$-reachable in }\\      			
			& S||G \text{ and } \mathcal{A}||G \text{ resp.})   \}\nonumber
		\end{align*}		
		\noindent	is a  simulation relation from $S||G$ to $\mathcal{A}||G$. 
		
		First, it will be verified that $\Psi$ satisfies the first condition in Definition \ref{def-conventional simulation}.		
		Let $(y_0,x_0)\in Y_0\times X_0$ and $x_0'\in X_0$. Since $\Phi$ is a simulation relation from $S||G$ to $R$ and $(y_0,x_0')\in Y_0\times X_0$, there exists $z_0'\in Z_0$ such that $((y_0,x_0'),z_0')\in \Phi$. Then, it follows from Lemma \ref{lemma-couter-touying is  simulation} that $(x_0',z_0')\in \pi(\Phi)$ and $\pi(\Phi): G\sqsubseteq_{uc} R$. Thus, by clause $(\ref{prop-fixpoint}-a)$ in Proposition \ref{prop-fixpoint}, $\pi(\Phi)=F_{(G,R)}(\pi(\Phi))$, and hence $(x_0',z_0')\in \pi(\Phi)\subseteq W_{(G,R)}^\uparrow$. 
		
		We have shown that, for any $x_0'\in X_0$, there exists $z_0'\in Z_0$ such that $(x_0',z_0')\in W_{(G,R)}^\uparrow$ and $((y_0,x_0'),z_0')\in \Phi$. Hence, for any $x_0'\in X_0$, we can choose arbitrarily and fix such a state  $z_0'$ and denote it as $\Delta(x_0')$. Set  $ W_0=\{(x_0',\Delta(x_0')): x_0'\in X_0\}.$
		Then $W_0\subseteq W_{(G,R)}^\uparrow\cap (X_0\times Z_0)$. Since $\mathcal{A}$ is a saturated $(G,R)$-automata, by clauses (state) and (sistate) in Definition \ref{def-S arrow new}, we have $W_0\in A_0$, moreover,  $((y_0,x_0'),z_0')\in \Phi$ for any $(x_0',z_0')\in W_0$ due to the definition of $W_0$. Thus, $((y_0,x_0),(W_0,x_0))\in \Psi$. Further, since $(W_0,x_0)$ is an initial state of $\mathcal{A}||G$, the first condition in  Definition \ref{def-conventional simulation} holds.

		Next, we deal with  the second condition in Definition \ref{def-conventional simulation}.
		Let $((y,x),(W,x))\in \Psi$, $(y,x)\stackrel{\sigma}{\longrightarrow}(y^*,x^*)$ and $\sigma\in \Sigma$. Thus, both $(y,x)$ and $(W,x)$ are $s$-reachable for some $s\in \Sigma^*$. Hence,  $(y_0,x_{01})\stackrel{s}{\longrightarrow}(y,x)$  and $(W_0,x_{02})\stackrel{s}{\longrightarrow}(W,x)$ for some $y_0\in Y_0$, $x_{01}\in X_0$, $W_0\in A_0$ and $x_{02}\in X_0$.  To complete this proof, it suffices to show that $W\stackrel{\sigma}{\longrightarrow}W^*$ and $((y^*,x^*),(W^*,x^*))\in \Psi$ for some $W^*\in A$. To this end, we give the claim below first.
		
		\textbf{Claim} The clauses $(\ref{def-S arrow old}-a)$ and $(\ref{def-S arrow old}-b)$ in Definition \ref{def-S arrow old} hold  for $W$ and $\sigma$.
		
		Since $W_0\stackrel{s}{\longrightarrow}W$ and $x_{02}\stackrel{s}{\longrightarrow}x$ due to $(W_0,x_{02})\stackrel{s}{\longrightarrow}(W,x)$, by Lemma \ref{lemma-couter-pi(X)w2}, $x\in Reach(s,X_0)=\pi_G(W)$, and hence $(x,z)\in W$ for some $z\in Z$. Further, it follows from  $x\stackrel{\sigma}{\longrightarrow}x^*$ that clause $(\ref{def-S arrow old}-a)$ in Definition \ref{def-S arrow old} holds for $W$ and $\sigma$.

		In order to illustrate that clause $(\ref{def-S arrow old}-b)$ in Definition \ref{def-S arrow old} holds for $W$ and $\sigma$, let $(x_1,z_1)\in W$ and $x_1\stackrel{\sigma}{\longrightarrow}x_1'$. Thus, 
		$((y,x_1),z_1)\in \Phi$ because of $((y,x),(W,x))\in \Psi$ and $(x_1,z_1)\in W$. By Lemma \ref{lemma-couter-pi(X)w2}, it follows from $W_0\stackrel{s}{\longrightarrow}W$ and $x_1\in \pi_G(W)$ that $x_1$ is  $s$-reachable. Further, due to $y_0\stackrel{s}{\longrightarrow}y$, $(y,x_1)$ is also  $s$-reachable in  $S||G$,  moreover, $(y,x_1)\stackrel{\sigma}{\longrightarrow}(y^*,x_1')$ comes from $(y,x)\stackrel{\sigma}{\longrightarrow}(y^*,x^*)$ and $x_1\stackrel{\sigma}{\longrightarrow}x_1'$. Hence, it follows from $((y,x_1),z_1)\in \Phi$  that 
		$z_1\stackrel{\sigma}{\longrightarrow}z_1'$ and 	$((y^*,x_1'),z_1')\in \Phi$ for some $z_1'\in Z$. Then, since $(y^*,x_1')$ is  $s\sigma$-reachable in $S||G$, 	by  Lemma \ref{lemma-couter-touying is  simulation},  $(x_1',z_1')\in \pi(\Phi)$ and $\pi(\Phi)$ is a $\Sigma_{uc}$-simulation from $G$ to $R$. Further, by clause $(\ref{prop-fixpoint}-a)$ in Proposition \ref{prop-fixpoint}, we have $\pi(\Phi)=F_{(G,R)}(\pi(\Phi))$, and hence $(x_1',z_1')\in \pi(\Phi)\subseteq W_{(G,R)}^\uparrow$. Hence, clause $(\ref{def-S arrow old}-b)$ in Definition \ref{def-S arrow old} holds for $W$ and $\sigma$.
		
		Next, we turn to proving the theorem itself. By the proof of the above claim,  for any $(x_1,z_1)\in W$ and $x_1'\in X$ with $x_1\stackrel{\sigma}{\longrightarrow}x_1'$, there exists $z_1'\in Z$ such that  $z_1\stackrel{\sigma}{\longrightarrow}z_1'$, $(x_1',z_1')\in W_{(G,R)}^\uparrow$ and 	$((y^*,x_1'),z_1')\in \Phi$. Thus, for each $(x_1,z_1)\in W$ and $x_1\stackrel{\sigma}{\longrightarrow}x_1'$, we can choose arbitrarily and fix such a state $z_1'$ and denote it as $\ast(x_1,z_1,x_1')$. Set 		 	
		$$W''= \{(x_1', \ast(x_1,z_1,x_1')): (x_1,z_1)\in W \text{ and } x_1\stackrel{\sigma}{\longrightarrow}x_1'  \}.       $$
		Clearly, $W''\in N(W,\sigma)$ and
		\begin{equation}\label{equation-counter theorem2}
			\ ((y^*,x_1'),z_1')\in \Phi\text{ for any }(x_1',z_1')\in W''.\tag{\ref{theorem-couter-maximall-permessive}-1}
		\end{equation}  
		
		\noindent	Since $\mathcal{A}$ is  a saturated $(G,R)$-automaton, by the above claim and clauses $(\ref{def-S arrow new}-a)$ and $(\ref{def-S arrow new}-c)$ in Definition \ref{def-S arrow new},  there exists  $W^*\in N(W,\sigma)$ such that $W\stackrel{\sigma}{\longrightarrow}W^*$ and $W^*\subseteq W''$, and hence $(W,x)\stackrel{\sigma}{\longrightarrow}(W^*,x^*)$ due to $x\stackrel{\sigma}{\longrightarrow}x^*$. Finally, we end the proof by showing $((y^*,x^*),(W^*,x^*))\in \Psi$.
		Since both $(W,x)$ and $(y,x)$ are $s$-reachable, 
		both $(y^*,x^*)$ and  $(W^*,x^*)$  are $s\sigma$-reachable due to $(W,x)\stackrel{\sigma}{\longrightarrow}(W^*,x^*)$ and $(y,x)\stackrel{\sigma}{\longrightarrow}(y^*,x^*)$.
		Moreover, $W^*$ satisfies $
		\ ((y^*,x_1'),z_1')\in \Phi \text{ for any }(x_1',z_1')\in W^*
		$   because of (\ref{equation-counter theorem2}) and $W^*\subseteq W''$. Hence,  $((y^*,x^*),(W^*,x^*))\in \Psi$, as desired.	    		
	\end{proof}     
\end{theorem}

The following two examples illustrate that the saturation conditions (i.e.\   clauses (sistate) and $(\ref{def-S arrow new}-c)$ in Definition \ref{def-S arrow new}) are necessary to ensure that $(G,R)$-automata are maximally permissive supervisors.

\begin{example}
	Consider the plant $G$, specification $R$ and $(G,R)$-automata $\mathcal{A}$ and  $\mathcal{B}$
	given in Example \ref{example-class example}. The automaton $\mathcal{A}$ satisfies one of the saturation conditions, i.e., clause (sistate) in Definition \ref{def-S arrow new}, but it doesn't satisfy clause $(\ref{def-S arrow new}-c)$.
	The reachable parts of $\mathcal{A}||G$ and $\mathcal{B}||G$ are presented graphically in Figure \ref{class example3}. It can be checked straightforwardly that  $\mathcal{A}$, $\mathcal{B}\in \SP(G,R)$.  However, since the state $(W_3,x_2)$ which can perform a $c$-labeled transition in $\mathcal{B}||G$ can't be matched by any state in $\mathcal{A}||G$, we get $\mathcal{B}||G\not\sqsubseteq\mathcal{A}||G$. 
	
	\begin{figure}[htbp]\vspace{-2ex}\centering
		\includegraphics[scale=0.6]{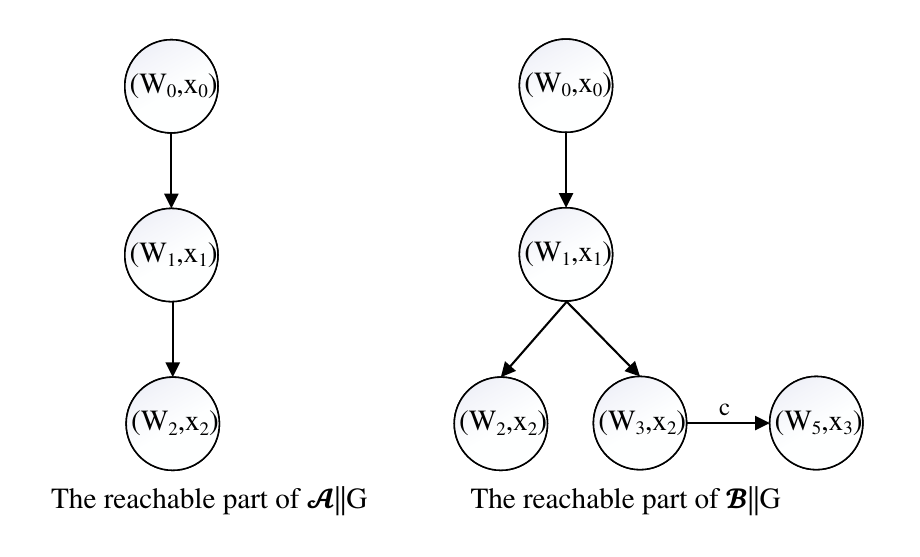}
		\caption{Necessity of the condition $(\ref{def-S arrow new}-c)$}
		\label{class example3}
	\end{figure}
\end{example}

\vspace{-4ex}

\begin{example}
	Consider the plant $G_1$, supervisor $S_1$ and specification $R_1$ given in Figure \ref{class example-sistate},  where $c_1,c_2\in \Sigma_{c}$ and all omitted labels are $\sigma\in \Sigma_{uc}$. Clearly, $G_1\sqsubseteq_{uc}R_1$, $S_1||G_1\sqsubseteq R_1$ and 	$
	W_{(G_1,R_1)}^\uparrow=\{(x_0,z_{01}),(x_0,z_{02})\}\cup (\{x_i:1\leqslant i\leqslant3\}\linebreak\times
	(\{z_i:1\leqslant i \leqslant 4\}\cup\{z_{01},z_{02}\}))$.	 
	For the automaton $\mathcal{A}_1=(A,\Sigma,\longrightarrow,A_0)$ with the reachable part shown in Figure \ref{class example-sistate}, $A=\powerset(W_{(G_1,R_1)}^\uparrow)$ and 	$A_0=\{W_{0}\}$ with $W_0=\{(x_0,z_{01})\}$, $W_1=\{(x_1,z_1)\}$, $W_2=\{(x_2,z_3)\}$.
	\begin{figure}[ht]\centering
		\vspace{-0.2cm}  
		\includegraphics[scale=0.58]{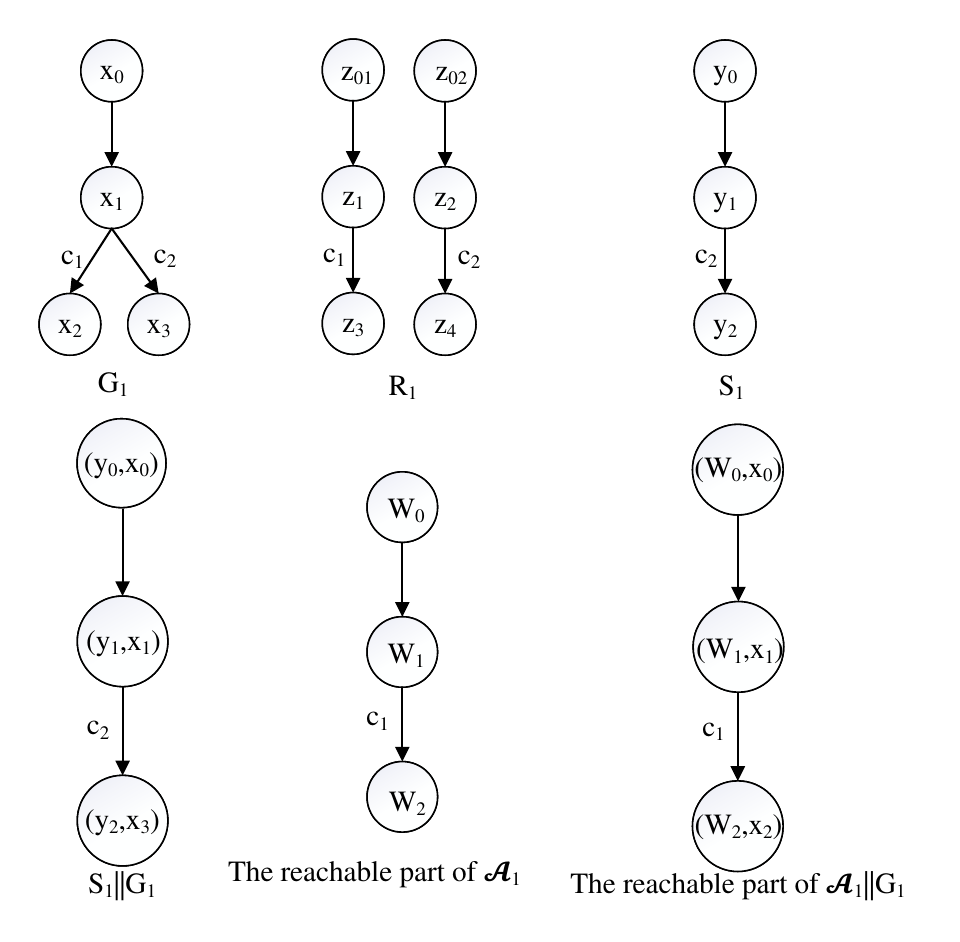}
		\caption{Necessity of the condition (sistate)}
		\label{class example-sistate}		
	\end{figure}
        It can be checked straightforwardly that $\mathcal{A}_1$ is a $(G_1,R_1)$-automaton and satisfies one of the saturation conditions, i.e., clause $(\ref{def-S arrow new}-c)$ in Definition \ref{def-S arrow new}. However, it isn't saturated because it doesn't satisfy clause (sistate) due to $\{(x_0,z_{02})\}\notin A_0$. Clearly, $\mathcal{A}_1\in \SP(G_1,R_1)$ and $S_1||G_1\not\sqsubseteq \mathcal{A}_1||G_1$.
        
Consider an automaton $\mathcal{B}=(B,\Sigma,\longrightarrow,B_0)$ shown in Figure \ref{figure-motivating example3} with $B=\powerset(W_{(G_1,R_1)}^\uparrow)$, $B_0=\{W_{01},W_{02}\}$, $W_{01}=\{(x_0,z_{01})\}$, $W_{02}=\{(x_0,z_{02})\}$, $W_3=\{(x_1,z_2)\}$ and $W_4=\{(x_3,z_4)\}$. Then it satisfies clause (sistate). Moreover, by Definition \ref{def-S arrow new}, it can be verified that $\mathcal{B}$ is a saturated automaton. Then it follows from Theorem \ref{theorem-couter-maximall-permessive} that $\mathcal{B}$ is a maximally permissive supervisor.

\begin{figure}[h]
	\centering
	\begin{minipage}[t]{0.4\textwidth}
    \centering
    \begin{tikzpicture}[
        ->, >=stealth,
        every node/.style={align=center}, 
        xnode/.style={
            draw, 
            circle, 
            minimum size=0.6cm,
            inner sep=0.6pt      
        },
        every path/.style={thick},
        font=\scriptsize
        ]
        
        \node[xnode] (W01) at (0, 0)    {$W_{01}$};
        \node[xnode] (W02) at (0, -1.2)  {$W_{02}$};
        \node[xnode] (W1)  at (1.2, 0)  {$W_{1}$};
        \node[xnode] (W2)  at (2.4, 0)  {$W_{2}$};      
        \node[xnode] (W3)  at (1.2, -1.2){$W_{3}$};
        \node[xnode] (W4)  at (2.4, -1.2){$W_{4}$};          
        
        \draw (W01) -- (W1);
        \draw (W1) -- (W2) node[midway, below]{c$_1$};
        \draw (W02) -- (W3);
        \draw (W3) -- (W4) node[midway, above]{c$_2$};

    \end{tikzpicture}
\end{minipage}
	\caption{The reachable part of $\mathcal{B}$}
	\label{figure-motivating example3}
\end{figure}

\end{example}

In the following, we intend to correct the flaws in \cite{2020Synthesis} based on Theorem \ref{th-couter-S arrow is a solution2} and \ref{theorem-couter-maximall-permessive}.
{Whereas example \ref{example-couter example} reveals that none of Theorem 1-3 obtained in \cite{2020Synthesis} is valid in general, below we will show that they do hold under the assumption that specifications are image-finite.  The key fact is that, under the assumption of image-finiteness, minimal elements always exist in the set $N(W,\sigma)$, which ensures  $\Sigma_{uc}$-admissibility and the saturation of the automaton $(G,R)^\uparrow$.}

	\vspace{-1ex}
	
	\begin{lemma}\label{lemma-exists minimal element in N(W)}
		Let $G$ and   $R$ be two automata and $(G,R)^\uparrow=(Y^\uparrow,\Sigma,\longrightarrow,Y_0^\uparrow)$. If $R$ is image-finite, then, for any $W\in Y^\uparrow$, $\sigma\in \Sigma$ and $W'\in    N(W,\sigma)$, there exists a minimal element $W^*$ in $N(W,\sigma)$ such that $W^*\subseteq W'$.

		\begin{proof}
			If $W'=\emptyset$, then $W'$ itself is minimal in $N(W,\sigma)$. In the following, we consider the case where $W'\neq\emptyset$.
			Put $\Omega=\{W''\in N(W,\sigma):W''\subseteq W'\}$ and let $\Omega'\subseteq \Omega$ be any chain under the relation $\subseteq$ (that is, $\Omega'\neq\emptyset$ and, for any $W,W'\in \Omega'$, either $W\subseteq W'$ or $W'\subseteq W$). By Zorn's lemma (see e.g.\  \cite{davey2002introduction}), if $\bigcap \Omega'\in \Omega$, then there exists a minimal element in the partially ordered set $\langle \Omega,\subseteq \rangle$, as desired. Next, we intend to show $\bigcap \Omega'\in \Omega$. Due to $\bigcap \Omega'\subseteq W'$, it is enough to prove $\bigcap \Omega' \in N(W,\sigma)$.
			
			Let $(x,z)\in W$ and $x^*\in X$ with $x\stackrel{\sigma}{\longrightarrow}x^*$. It suffices to show that $(x^*,z')\in \bigcap \Omega'$ for some $z'\in Z$ with $z\stackrel{\sigma}{\longrightarrow}z'$. On the contrary, we assume that there isn't such a state. Since $R$ is image-finite, there exist finitely many states in $Z$, say $z_0,z_1,...,z_n$, such that $z\stackrel{\sigma}{\longrightarrow}z_i$ $(0\leqslant i\leqslant n)$. By the assumption, for each $i\leqslant n$, $(x^*,z_i)\notin W_i$ for some $W_i\in \Omega'$. However, since $\Omega'$ is a chain, $\bigcap_{0\leqslant i \leqslant n}W_i=W_{i_0}$ for some $0\leqslant i_0\leqslant n$, and hence $\bigcap_{0\leqslant i \leqslant n}W_i\in N(W,\sigma )$, which contradicts that $(x,z)\in W$, $x\stackrel{\sigma}{\longrightarrow}x^*$ and $(x^*,z_i)\notin \bigcap_{0\leqslant i \leqslant n}W_i$ for each $0\leqslant i \leqslant n$.			
		\end{proof}
	\end{lemma}

	\vspace{-2ex}

	\begin{proposition}\label{proposition-S arrow is satureted}
		Given a plant $G$ and   specification $R$ with $G\sqsubseteq_{uc}R$, if $R$ is   image-finite,  	then $(G,R)^\uparrow$ is a  
		saturated $(G,R)$-automaton.
		
		\begin{proof}
			Let $(G,R)^\uparrow=(Y^\uparrow,\Sigma,\longrightarrow,Y_0^\uparrow)$. It immediately follows from the fact below that the automaton
			$(G,R)^\uparrow$ satisfies clause (sistate) in Definition \ref{def-S arrow new}.		
			\begin{align*}
				&\{W_0\in Y^\uparrow\cap\powerset(X_0\times Z_0):\forall x_0\in X_0\  \exists !z_0\in Z_0((x_0,z_0)\\[-1ex]
				&\in W_0)\}\subseteq\{W_0\in Y^\uparrow\cap\powerset(X_0\times Z_0): |W_0|=|X_0| \text{ and } \\[-1ex]
				&\forall x_0\in X_0 \ \exists z_0\in Z_0((x_0,z_0)\in W_0)\}=Y_0^\uparrow.
			\end{align*}	
			\noindent 	Moreover, it satisfies all other clauses in Definition \ref{def-S arrow new} trivially except  	$(\ref{def-S arrow new}-a)$ and
			$(\ref{def-S arrow new}-c)$. Next we verify that $(G,R)^\uparrow$ satisfies them in turn.

			$(\ref{def-S arrow new}-a)$  Let $W\in Y^\uparrow$ and $\sigma\in \Sigma$. Assume that $W$ and $\sigma$ satisfy clauses $(\ref{def-S arrow old}-a)$ and
			$(\ref{def-S arrow old}-b)$ in Definition $\ref{def-S arrow old}$. Then, $W\neq\emptyset$ due to $(\ref{def-S arrow old}-a)$. Put $W'=\{(x',z'):\exists(x,z)\in W
				(x\stackrel{\sigma}{\longrightarrow}x', z\stackrel{\sigma}{\longrightarrow}z'  \text{ and } (x',z')\in W_{(G,R)}^\uparrow)\}.$         	 Thus, $ W'\subseteq W_{(G,R)}^\uparrow$. Moreover, since $W$ and $\sigma$ satisfy clauses $(\ref{def-S arrow old}-b)$ and $G\sqsubseteq_{uc}R$, by Observation \ref{obs-1}, for any 	$ (x,z)\in W$   and $x'\in X$ with  $x\stackrel{\sigma}{\longrightarrow} x'$, there exists $z'\in Z$  such that $z\stackrel{\sigma}{\longrightarrow}z'$ and $(x',z')\in W_{(G,R)}^\uparrow$. Hence, $ W'\in N(W,\sigma) $. Then, by Lemma \ref{lemma-exists minimal element in N(W)}, there exists a minimal element $W^*$ in $N(W,\sigma)$ such that $W^*\subseteq W'$. Further, by Definition \ref{def-S arrow old}, we have $W\stackrel{\sigma}{\longrightarrow}W^*$, as desired.
			
			$(\ref{def-S arrow new}-c)$ Assume $W\in Y^\uparrow$ and $\sigma\in \Sigma$ with $W\stackrel{\sigma}{\longrightarrow}$. Let $W'\in N(W,\sigma)$. Due to $W\stackrel{\sigma}{\longrightarrow}$, by Definition \ref{def-S arrow old}, clauses 	$(\ref{def-S arrow old}-a)$ and 	$(\ref{def-S arrow old}-b)$ in  Definition \ref{def-S arrow old} hold for $W$ and $\sigma$. Moreover, by Lemma \ref{lemma-exists minimal element in N(W)}, there exists a minimal element $W^*$ in $N(W,\sigma)$ such that $W^*\subseteq W'$. Thus, by Definition \ref{def-S arrow old}, $W\stackrel{\sigma}{\longrightarrow}W^*$, as desired.
		\end{proof}
	\end{proposition}

	By  the above proposition and Theorem \ref{theorem-couter-maximall-permessive}, we could  obtain the  following corollaries, which show that the methods for synthesizing maximally permissive supervisors 
	proposed in \cite{2020Synthesis} indeed work well under the assumption of image-finiteness.

	\vspace{-1ex}
	\begin{corollary}(Correction of Theorem 1 and 2 in \cite{2020Synthesis})\label{corollary-1}
		For any plant $G$ and  specification $R$ with $G \sqsubseteq_{uc} R$,  if $R$ is image-finite, then $(G,R)^\uparrow\in \MSP(G,R)$.
	\end{corollary}
	
	\vspace{-1ex}
	For the automaton $\widetilde{(G,R)^\uparrow}$, we also have the following result. Its proof is the same as that in \cite{2020Synthesis} except  using Corollary \ref{corollary-1} above instead of Theorem 1 and 2  in \cite{2020Synthesis}.
	\vspace{-1ex}

	\begin{corollary}(Correction of Theorem 3 in \cite{2020Synthesis}) \label{corollary-2}
		Let $G$ and  $R$ be two automata with $G \sqsubseteq_{uc} R$. If $R$ is 
		image-finite, then  $\widetilde{(G,R)^\uparrow}\in \MSP(G,R)$.
	\end{corollary}

	We can modify clause $(\ref{def-S arrow old}-c)$ in Definition $\ref{def-S arrow old}$ to provide methods of constructing maximally permissive supervisors that solve the $(G,R)$-similarity control problem whenever $G \sqsubseteq_{uc} R$, which don't depend on the assumption of image-finiteness. In other words, we can modify Takai's construction to make Theorem 1-3 in \cite{2020Synthesis} hold in general.
	In detail, we have the following two corollaries. It's not difficult to verify that  both of the below modifications  will result in saturated $(G,R)$-automata. Therefore, by Theorem \ref{th-couter-S arrow is a solution2} and \ref{theorem-couter-maximall-permessive}, they hold immediately.

	\begin{corollary}(Modify Takai's Construction 1)\label{corollary-construction1}
		For any automaton $G$ and $R$, if $G\sqsubseteq_{uc}R$, then $(G,R)_1^\uparrow\in \MSP(G,R)$, where $(G,R)_1^\uparrow$ is defined as Definition $\ref{def-S arrow old}$ except that clause $(\ref{def-S arrow old}-c)$ is weakened as
		
		$(\ref{def-S arrow old}-c') W'\in N(W,\sigma)$.	
	\end{corollary}
	
	\vspace{-3ex}
	
	\begin{corollary}(Modify Takai's Construction 2)\label{corollary-construction2}
		For any automaton $G$ and $R$, if $G\sqsubseteq_{uc}R$, then $(G,R)_2^\uparrow\in \MSP(G,R)$, where $(G,R)_2^\uparrow$ is defined as Definition $\ref{def-S arrow old}$ except that clause $(\ref{def-S arrow old}-c)$ is weakened as
		
		$(\ref{def-S arrow old}-c'')$  $W'$ satisfies one of the following conditions:
		
		\ \ $(\ref{def-S arrow old}-c''-1)$  $W'$ is minimal in $N(W,\sigma)$;
		
		\ \ $(\ref{def-S arrow old}-c''-2)$  $W'\in N(W
		,\sigma)$ and there is no minimal element in the set $\{W''\in N(W,\sigma): W''\subseteq W'\}.$
	\end{corollary}
	
	In order to meet the condition $(\ref{def-S arrow new}-c)$ in Definition \ref{def-S arrow new}, compared to clause $(\ref{def-S arrow old}-c)$, $(\ref{def-S arrow old}-c')$ simply removes the requirement for successors to be minimal elements. One advantage of $(\ref{def-S arrow old}-c'')$ over $(\ref{def-S arrow old}-c')$ is that it may reduce the number of reachable states in some situations (e.g., $N(W,\sigma)$ contains no infinite decreasing chains.) Here, we intend to provide theoretical results about a maximally permissive supervisor in general cases rather than its effective constructing method\footnote{For finite automata, such an effective constructing method has been given by Takai in \cite{2020Synthesis}, see Definition \ref{def-S arrow old} in this paper.}, so these results do not rely on the decidability of predicates like ``$W'$ is minimal in $N(W,\sigma)$" and ``there is no minimal element in the set $\{W''\in N(W,\sigma): W''\subseteq W'\}$".
	
	Moreover, in order to further reduce the number of reachable states, we can define the set of initial states of $(G,R)^\uparrow$ and $(G,R)_i^\uparrow$ $(i=1,2)$ as
	\begin{align*}
		\{&W_0\in Y^\uparrow\cap\powerset(X_0\times Z_0):\forall x_0\in X_0\\\ &\exists! z_0\in Z_0((x_0,z_0)\in W_0)\},
	\end{align*} 
	which is a subset of 
	\begin{align*}
		\{&W_0\in Y^\uparrow\cap\powerset(X_0\times Z_0):|W_0|=|X_0| \text{ and }\\&\forall x_0\in X_0 \ \exists z_0\in Z_0((x_0,z_0)\in W_0)\}=Y_0^\uparrow.
	\end{align*}
	
	To apply the construction of \cite{2020Synthesis} in non-image-finite systems, one may transform them to image-finite ones at the cost of introducing infinitely many initial states, as mentioned in Section 2.  However, this transformation can be circumvented by modifying the construction of \cite{2020Synthesis} in accordance with  Corollary \ref{corollary-construction1} or \ref{corollary-construction2}.

	By the way,  the following implication doesn't always  hold, where $W_0\in Y_0^\uparrow$ in $(G,R)^\uparrow$,
	\begin{align*}
		& |W_0|=|X_0| \text{ implies } \forall (x_0,z_0),(x_0',z_0')\in W_0 \\
		&(x_0=x_0'\Longrightarrow (x_0,z_0)=(x_0',z_0')).
	\end{align*}
	\noindent For instance, consider  $W_0$ given in Example \ref{example-couter example}. 
	We end this section by providing examples of $(G,R)_i^\uparrow$ ($i=1,2$).

	\vspace{-1ex}
	\begin{example}
		Consider the plant $G$ and specification $R$ from Example \ref{example-couter example}. Let  $W_{01}=\{(x_{0i},z_{0i}):i\geqslant1\}\cup\{(x_0,z_{01}), (x_0,z_{02})\}$ and $W_1=\{(x,z_1),(x,z_2)\}$. Clearly,  $W_{01}\in Y_0^\uparrow$, $W_1\in Y^\uparrow$ and $W_{01}\stackrel{\sigma}{\longrightarrow}W_1$ in $(G,R)^\uparrow$, $(G,R)_1^\uparrow$ and $(G,R)_2^\uparrow$. Since $N(W_1,\sigma)=\powerset(\{x'\}\times\{z_i':i\geqslant1\})-\{\emptyset, \{(x',z_1')\}\}$ and the set $T=\{\{(x',z_i')\}:i\geqslant 2\}$ consists of all the minimal elements in $N(W_1,\sigma)$,  for any $W_1'\in N(W_1,\sigma)$, we have $W_1\stackrel{\sigma}{\longrightarrow}W_1'$ in $(G,R)_1^\uparrow$ and
		 	$W_1\stackrel{\sigma}{\longrightarrow}W_1'$\text{ in }$(G,R)_2^\uparrow\  (\text{or, }  (G,R)^\uparrow)  \text{ iff } W_1'\in T.$               
		
		For the state  $W$ in  Example \ref{example-couter example}, it's a deadlock in $(G,R)^\uparrow$, while $W\stackrel{\sigma}{\longrightarrow}W'$ in both $(G,R)_1^\uparrow$ and $(G,R)_2^\uparrow$ for any  infinite subset $W'$ of $\{(x',z_i'):i\geqslant 1\}$. Therefore, $(G,R)^\uparrow$, $(G,R)_1^\uparrow$ and $(G,R)_2^\uparrow $ are different from each other.
	\end{example}

	\noindent

	\section[Flaws in Li \& Takai 2019/2022]{Flaws in \cite{2022Synthesis,9029606}}\label{sec:flaws}
	
	The method for synthesizing a maximally permissive supervisor that solves the similarity control problem, due to Takai, has been extended to handle the corresponding problems under partial observation in \cite{2022Synthesis,9029606}. Unfortunately, similar flaws also occur in these works. This section intends to illustrate this briefly.

	In addition to dividing events into controllable and uncontrollable ones as in \cite{2020Synthesis}, since it is assumed that the occurrence of events is partially observable, the set $\Sigma$ is further partitioned into observable event set $\Sigma_{o}$ and unobservable event set $\Sigma_{uo}$ in \cite{2022Synthesis,9029606}, that is, $\Sigma=\Sigma_{uc}\cup \Sigma_{c}=\Sigma_{uo}\cup \Sigma_{o}$. Moreover, the set   $\Gamma_{uo}$ is defined as $\Gamma_{uo}=\{\gamma\in \powerset(\Sigma_{uo}):\Sigma_{uc}\cap \Sigma_{uo}\subseteq\gamma \}$.
	
	In order to capture the intuition that the supervisor does not react to the occurrence of any unobservable event,
	the notion of $\Sigma_{uc}$-admissibility is strengthened in \cite{2022Synthesis,9029606} and recalled below.
	\smallskip
	
	\begin{myDef}\label{def-admissible under partial }
		Given a plant $G=(X,\Sigma, \longrightarrow, X_0)$, a supervisor $S=(Y,\Sigma, \longrightarrow, Y_0)$ is said to be admissible with respect to $G$ under partial observation if $S$ is $\Sigma_{uc}$-admissible with respect to $G$ and, for any reachable state $(y,x)$ in $S||G$, $\sigma\in \Sigma_{uo}$  and $y_1\in Y$,
		$$y\stackrel{\sigma}{\longrightarrow}y_1 \text{ implies } y_1=y. $$
	\end{myDef}

	\vspace{-2ex}
	
	Given the automata $G$ and $R$, the so called \textit{$(G,R)$-similarity control problem  under partial observation} refers to  finding a  supervisor $S$ which is  admissible (w.r.t.\  $G$) under partial observation such that $S||G\sqsubseteq R$ \cite{2022Synthesis,9029606}.
	Next we recall the method, proposed in  \cite{2022Synthesis,9029606}, for constructing a maximally permissive supervisor.
	\smallskip

	\begin{myDef}\label{def-other mistake1}
		Let $G=(X,\Sigma, \longrightarrow, X_{0})$ and $R=(Z,\Sigma, \longrightarrow,[2] Z_{0})$  be two  automata. The state set $Y^\Uparrow \subseteq \powerset(W_{(G,R)}^\uparrow)\times \Gamma_{uo} \times \powerset(W_{(G,R)}^\uparrow)$ is defined as 
		\begin{align*}
			&Y^\Uparrow=\{ (W_1,\gamma_{uo},W_2): W_1\neq\emptyset,  W_2\text{ is minimal in }\\
			&U(W_1, \gamma_{uo}) \text{ and }\forall \sigma\in  \gamma_{uo}\cap\Sigma_{c}\ \exists(x,z)\in W_2(x\stackrel{\sigma}{\longrightarrow}) 
			\}, 
		\end{align*}
		where $U(W_1, \gamma_{uo})=$ 
		\begin{align*}
			\{&W\in \powerset(W_{(G,R)}^\uparrow):   W_1\subseteq W \text{ and }\forall \sigma\in \gamma_{uo} \ \forall (x,z)\in W\\\ 
			&\forall x' (x\stackrel{\sigma}{\longrightarrow} x'\Longrightarrow \exists z'(z\stackrel{\sigma}{\longrightarrow}z'\text{ and }  (x',z')\in W))   \}.
		\end{align*}
		The initial state set $Y_0^\Uparrow\subseteq Y^\Uparrow$ is defined as 
		\begin{align*}
			Y_0^\Uparrow=\{&(W_{01},\gamma_{uo},W_{02})\in Y^\Uparrow:  |W_{01}|=|X_0| \text{ and }\\
            &\forall x_0\in X_0\ \exists z_0\in Z_0((x_0,z_0)\in W_{01})\}.
		\end{align*}
		
	\end{myDef}

	\vspace{-3ex}
	
	\begin{myDef}
		\label{def-other mistake2}
		Let $G=(X,\Sigma, \longrightarrow, X_{0})$ and $R=(Z,\Sigma, \longrightarrow,\linebreak  Z_{0})$  be two  automata. A supervisor $(G,R)^\Uparrow$ is defined as $(G,R)^\Uparrow=(Y^\Uparrow,\Sigma,\longrightarrow, Y_0^\Uparrow)$, where, 	
		for any $y=(W_1,\gamma_{uo},W_2),y'=(W_1',\gamma_{uo}',W_2')\in Y^\Uparrow$ and $\sigma\in \Sigma$, $y\stackrel{\sigma}{\longrightarrow}y'$ iff
		
		$(\ref{def-other mistake2}-a)$ either	$\sigma\in \gamma_{uo}$ and  $y=y'$, or

		$(\ref{def-other mistake2}-b)$ $\sigma\in \Sigma(y)$ and $W_1
		'$ is minimal in $M(W_2,\sigma)$ with 
		\begin{align*}
			&M(W_2,\sigma)=\{W''\in \powerset(W_{(G,R)}^\uparrow)
			: \forall (x,z)\in W_2\  \forall x'\\&(x\stackrel{\sigma}{\longrightarrow} x'\Longrightarrow \exists z'(z\stackrel{\sigma}{\longrightarrow}z'\text{ and }  (x',z')\in W''))\},
		\end{align*}

		$\Sigma(y)=\{\sigma\in \Sigma_{o}: \exists (x,z)\in W_2(x\stackrel{\sigma}{\longrightarrow})  \text{ and } \forall (x,z)\in W_2\  \forall x'(x\stackrel{\sigma}{\longrightarrow} x'\Longrightarrow \exists z'(z\stackrel{\sigma}{\longrightarrow}z'\text{ and }  (x',z')\in W_{(G,R)}^\uparrow))\}$.

	\end{myDef}

	In the situation that $\Sigma_{uo}=\emptyset$, since the relevant conditions involved in Definition \ref{def-other mistake1} and \ref{def-other mistake2} are true trivially, the constructions given in Definition \ref{def-S arrow old} and \ref{def-other mistake2} respectively are the same in essence. In detail, let  $G=(X,\Sigma, \longrightarrow, X_{0})$ and $R=(Z,\Sigma, \longrightarrow, Z_{0})$  be two  automata with $G\sqsubseteq_{uc} R$. Assume that $(G,R)^\uparrow=(Y^\uparrow,\Sigma,\longrightarrow, Y_{0}^\uparrow)$ and $(G,R)^\Uparrow=(Y^\Uparrow,\Sigma,\longrightarrow, Y_{0}^\Uparrow)$ are automata defined by  Definition \ref{def-S arrow old} and \ref{def-other mistake2} respectively. By Definition \ref{def-S arrow old}, \ref{def-admissible under partial }, \ref{def-other mistake1} and \ref{def-other mistake2}, the following facts can be checked straightforwardly whenever $\Sigma_{uo}=\emptyset$.
	
	(0) For any automaton $S$, $S$ is admissible in the sense of Definition \ref{def-admissible under partial 
	} iff $S$ is $\Sigma_{uc}$-admissible.
	
	(1) $Y^\Uparrow=\{(W,\emptyset,W):W\in Y^\uparrow \text{ and } W\neq\emptyset \}$;
	
	(2) $Y_{0}^\Uparrow=\{(W,\emptyset,W):W\in Y_{0}^\uparrow\}$;
	
	(3) For any $W$, $W'\in Y^\uparrow-\{\emptyset\}$ and $\sigma\in \Sigma$,
	$(W,\emptyset,W)\stackrel{\sigma}{\longrightarrow}(W',\emptyset,W') \text{ in } (G,R)^\Uparrow\text{ iff } W\stackrel{\sigma}{\longrightarrow}W' \text{ in } (G,R)^\uparrow;$
	
	Hence, $(G,R)^\uparrow$ and $(G,R)^\Uparrow$ are essentially the same except $\emptyset\in Y^\uparrow$ but $(\emptyset,\emptyset,\emptyset)\notin Y^\Uparrow$. Further,  since the state $\emptyset$ is isolated and  not an initial state in $(G,R)^\uparrow$, we also have
	
	(4) $(G,R)^\uparrow\in \SP(G,R)$ if $(G,R)^\Uparrow$ can solve the $(G,R)$-similarity control problem under partial observation;

	(5) $(G,R)^\uparrow\in \MSP(G,R)$ if $(G,R)^\Uparrow$ is a maximally permissive supervisor for the $(G,R)$-similarity control problem under partial observation.

	By the facts (4) and (5) above,  Example \ref{example-couter example} can also act as an counterexample for Theorem 1 and 
	2 in  \cite{9029606}  and Theorem 18 and 24 in \cite{2022Synthesis}    with $\Sigma_{uo}=\emptyset$, which assert that, for any $G$ and $R$ with $G\sqsubseteq_{uc}R$, $(G,R)^\Uparrow$ is a maximally permissive supervisor for the $(G,R)$-similarity control problem under partial observation. Consequently, none of 
	these theorems  holds in general.
	
	\section{Conclusion}\label{sec:conc}
	Taking inspiration from Takai's construction (i.e.\  Definition \ref{def-S arrow old}), this paper introduces the notion of a saturated $(G,R)$-automaton. It has been demonstrated that, for any automata $G$ and $R$ with $G\sqsubseteq_{uc}R$, each $(G,R)$-automaton is a solution of the $(G,R)$-similarity control problem, while saturated ones are maximally permissive supervisors. Our work reveals that saturatedness plays a central role in constructing maximally permissive supervisors in the style of Definition \ref{def-S arrow old}. Since minimal elements do not always exist in a set like $N(W,\sigma)$, the automaton $(G,R)^\uparrow$ is neither saturated nor $\Sigma_{uc}$-admissible in general. Thus, the methods due to Takai (i.e.\  Definition \ref{def-S arrow old}) are not applicable to general situations. However, Corollary \ref{corollary-1} and \ref{corollary-2} indicate that they indeed work well in the situation where the specification $R$ is image-finite.

This paper provides two approaches to decide whether a supervisor $S\in\SP(G,R)$ is a maximally permissive solution. The first is  checking whether $S$ satisfies the saturation condition. If it does, then by Theorem \ref{theorem-couter-maximall-permessive}, $S$ is maximally permissive. One of the shortcomings of this approach lies in the fact that the notion of saturated automata depends on the representation of states, which restricts the development of methods for deciding the maximality of solutions based on this theorem.  For the second approach, the automaton $(G,R)^\uparrow$  should be constructed first, based on Corollary \ref{corollary-construction1}. Then whether $S$ is maximally permissive can be decided through checking whether $(G,R)^\uparrow||G\sqsubseteq S||G$. This follows from the below observation:
$$
S\in \MSP(G,R) \text{ iff } (G,R)^\uparrow||G\sqsubseteq S||G.$$
\noindent Compared with the first method, the advantage of the latter is that it provides a sufficient and necessary determining approach, while the former is only sufficient. However, the complexity of constructing $(G,R)^\uparrow$ is exponential, which is the same as verifying saturation, while the second method incurs more overhead due to the additional step of judging the simulation relation.

The construction of a maximally permissive solution through the method of Corollary \ref{corollary-construction1}
   may involve representing the infinitely many states by finite means, for instance using words over an alphabet to denote states or employing finite-state machines with an additional  component like a stack or tape, such as pushdown automata. Transitions are then defined by finite-state devices or rational rewriting rules that manipulate these state representations. Hence, from a computational perspective, the decision procedures presented in this paper are only applicable to finite automata until these issues are resolved.
   We leave them for future work.

We conclude this paper with the  following  unresolved conjecture raised by an anonymous reviewer.

\textit{Conjecture}: Let $G$ and $R$ be two automata. If $G\sqsubseteq_{uc}R$ and $G$ is finite, then $(G,R)^\uparrow\in \MSP(G,R)$.

	\vspace{-1ex}
    \section*{References}
        \vspace{-2em}
    \bibliographystyle{eptcs}
	\bibliography{Reference}
	
\end{document}